\documentclass[preprintnumbers,article,amsmath,amssymb,floatfix,10pt,prd,twocolumn,
superscriptaddress,nofootinbib]{revtex4}
\usepackage[colorlinks=true, pdfstartview=FitV, linkcolor=blue, citecolor=red, urlcolor=magenta]{hyperref}
\usepackage{bbm}
\usepackage{amsfonts}
\usepackage{mathrsfs}
\usepackage{latexsym}
\usepackage{epsfig}
\usepackage{epstopdf}
\usepackage{epstopdf}
\usepackage{graphicx}
\usepackage{amssymb}
\usepackage{amsmath}
\usepackage{dcolumn}
\usepackage{bm}
\usepackage{float}
\usepackage{color}
\usepackage{comment}
\usepackage{xcolor}
\begin{document}

\title{\bf Dynamical stability of new wormhole solutions via cold dark matter and
solitonic quantum wave halos in $f(\mathcal{R},\mathcal{L}_m)$ gravity}
\author{G. Mustafa}
\email{gmustafa3828@gmail.com}\affiliation{Department of Physics, Zhejiang Normal University, Jinhua 321004, People's Republic of China}

\author{Faisal Javed}
\email{faisaljaved.math@gmail.com}\affiliation{Department of
Physics, Zhejiang Normal University, Jinhua 321004, People's
Republic of China}

\author{S.K. Maurya}
\email{sunil@unizwa.edu.om}\affiliation{Department of Mathematical and Physical Sciences, College of Arts and Sciences, University of Nizwa, Nizwa 616, Sultanate of Oman}

\author{M. Govender}
\email{megandhreng@dut.ac.za}\affiliation{Department of Mathematics, Durban University of Technology, Durban 4000, South Africa}

\author{Amna Saleem}
\email{a.saleemintel@gmail.com}\affiliation{Department of Mathematics, Riphah International University Faisalabad Campus, Islamabad Pakistan}

\date{\today}

\begin{abstract}
This current analysis offers novel wormhole solutions in the background of newly developed extended $f(\mathcal{R},\mathcal{L}_m)$ gravity. We use an anisotropic matter source and a particular type of energy density demonstrating cold dark matter halo and quantum wave dark matter halo to calculate two different wormhole solutions. The properties of the exotic matter within the wormhole geometry and the matter contents via energy conditions are studied in detail, both analytically and graphically, by showing valid and invalid regions. The calculated shape functions of wormhole geometry satisfy the required conditions in both cases. Further, we investigate the stability of the shell around wormhole structures by considering black hole solutions in the framework of cold dark matter halo and quantum wave dark matter by assuming matter contents located at the shell follow the phantom-like equations of state. Then,  for quintessence and phantom energy type equations of state, the choice of cold dark matter halo has maximum stability at lower equilibrium shell radii and declines as the radius rises. More petite wormhole throats have an unstable configuration for the dark energy matter content, while higher wormhole throat radius has the lowest stability. For the choice of quantum wave dark matter, the shell around the wormhole structure is unstable for both quintessence and phantom energy, while dark energy shows a stable configuration for smaller values of wormhole throat. 
\end{abstract}

\maketitle

\section{Introduction}
The term 'wormhole' was first appeared in a beautifully written paper entitled "Classical Physics as Geometry by Misner and Wheeler \cite{misnerwheeler}. The notion of wormholes was enunciated by Herman Weyl as far back as in his attempt to explain the topological nature of the electromagnetic field \cite{weyl1}. Einstein and Rosen, while attempting to model a particle within the framework of general relativity free of singularities, conjured up a picture of physical space as consisting of two congruent sheets where the particle is a portion of space connecting the two sheets. The connection is referred to as the bridge \cite{einsteinrosen}. On a larger scale, solutions of the Einstein field equations describe wormholes as theoretical structures that connect two distinct universes or two separated regions of the same universe by a throat region. The possibility of wormholes being used as portals connecting vast distances in the universe by providing ``shortcuts" in space-time has been romanticized by the layperson and has entered the realm of science fiction. 

Observations of gravitational waves and black hole shadows have cast new light onto the existence of superdense, gargantuan vacuum machines, which we refer to as black holes. Supermassive black holes with masses ranging from $10^6$ to $10^10$ $M_{\odot}$ are known to exist at the center of nearly every galaxy. Early observations by the Event Horizon Telescope collaboration (EHT) of the supermassive black hole at the heart of the elliptical galaxy M87 estimate its mass to be of the order of $10^9$ $M_{\odot}$\cite{EHT1, EHT2,EHT3}. In so far as gravitational wave events are concerned, the GW190814 event (LIGO-Virgo collaboration) points to the signals giving rise to this observation as results of the binary coalescence of a $23.2^{+1.1}_{-1.0}{M_\odot}$ black hole (BH) and a compact object with a mass of 2.50 to 2.67$M_{\odot}$\cite{gw}. The secondary component of this merger presents a peculiar conundrum for the nature of the secondary component of GW190814 as either the lightest black hole or the heaviest neutron star ever observed in a binary system. Studies have proposed that the wormhole masquerades as a black hole to a faraway observer for large gravitational redshift at the throat. In this scenario, the throat mimics the event horizon, a uni-directional causal membrane. 

The idea of wormholes as being conduits for interstellar travel has prompted a plethora of investigations into their nature and structure. The first wormhole solution referred to in the literature as the Schwarzschild wormhole or the Einstein-Rosen bridge \cite{einsteinrosen} borne out of classical general relativity is non-traversable, i.e., material objects cannot tunnel through. Solutions of the Einstein field equations describing traversable wormholes followed soon \cite{mt1}. The caveat here is that these solutions violate the standard energy conditions. To circumnavigate this pathology, one is required to invoke exotic matter content. Visser and co-workers have played a key role in studying wormhole solutions in traversability, stability, energy conservation, and wormholes in alternative theories of gravity, viz., braneworld scenarios. Wormhole solutions were also utilized to study gravitational lensing. In the strong lensing regime, Shaikh et al. \cite{shaikh} showed that the location of the observer (i. observer and the source are on the same side of the wormhole throat, ii. observer and the source are on opposite sides of the throat) plays a fundamental role in distinguishing between black hole and wormhole backgrounds. 

Finding a convincing and reliable explanation for the noteworthy events has become imperative in light of the recent finding of the cosmic aspects, which were first confirmed by measurements in \cite{njim1,njim2}. This also brought up the question of the strange composition. Thus, it is necessary to investigate more general extensions of the standard general theory of relativity. The Generalized theories of gravity models incorporating relationships between space-time geometry and matter existence are of great interest. These investigations have discussed numerous significant phenomena, like late-time cosmic acceleration, integrating inflationary models with dark energy, and eliminating particular dark matter possibilities by analyzing enormous test particles. Reasonable explanations for these events within the context of $f(\mathcal{R})$ theory of gravity are given in Ref. \cite{njim3,njim4}.
The geometry theories modify the gravitational Lagrangian by taking into account random functions of geometric components. This approach is extended by linking the geometry and matter content of the universe by integrating the matter Lagrangian in addition to geometric description. Several scientific aspects of such curvature–matter coupling theories have been addressed by different authors in \cite{rjim1,rjim2,rjim3,rjim4,rjim5}. A reasonable work has been done by Capozziello et al. \cite{njim10,njim11,njim12,njim13,njim14,njim15,njim16,njim17} in different theories of gravity for various aspects related to cosmos. The traversable wormhole solutions in the background of other modified gravity theories have been thoroughly explored in the literature \cite{bohmer2012,lobo2009,lobo2020,harko2013,kanti2011,usmani2010,rahaman2006,rahaman2012,zubair2016,ovgun2018,mustafa2021,cap2,cap3,cap4,cap5,kuhfittig2015,kuhfittig2005,ref1}. Recently, Kavya and Mustafa \cite{njim18,njim19,njim20} calculated new viable wormhole solutions in the background of $f(\mathcal{R},\mathcal{L}_m)$ gravity.

The wormhole solutions in the Rastall gravity framework are introduced in the paper \cite{ta10} by use of the phantom regime with conformal symmetry. Along with the wormhole structure, the authors examine the epicyclic frequencies and the stability of thin-shells surrounding it. The stability of thin-shells around new wormhole solutions in the setting of teleparallel gravity is examined by Javed et al. \cite{ta11}. Furthermore, \cite{ta12} describes in detail the investigation of wormhole solutions and thin-shell stability in the $f(R,T)$ gravity framework. In addition, the research in \cite{ta13} investigates traversable wormhole solutions in $f(Q)$ that exhibit twin peak quasi-periodic oscillations and checks if these structures are stable across thin-shell boundaries. The stability of thin-shells around these structures and the impact of quantum wave dark matter on wormhole solutions in General Relativity are also addressed in \cite{ta14}.

 Expanding on these earlier studies, the present research is devoted to calculate the wormhole solution in the framework of CDM halo and quantum wave dark matter distributions. Then, we are interested in developing the behavior of the shell formulated by considering a respected black hole solution filled with phantom-like EoS. The current work is organized as follows. In the next Sec, we explore the $f(\mathcal{R},\mathcal{L}_m)$ gravity and calculate its field equations. In the Sec. III, we provide the quantum dark matter halo and its background. Sec. IV deals with cold dark matter halo with NFW profile. In the Sec. V, we calculate the new wormhole solution under the effect of two different kinds of dark matter halos. In the next Sec. energy conditions and matter distribution are discussed. The second last Sec. explores the stability of the shells by using two different black hole solutions within the context of dark matter halos. In the last Sec. we summarize our results.
\section{$f(\mathcal{R},\mathcal{L}_m)$ Gravity and Wormhole Geometry}\label{II}

A generalized gravity version of  $f(\mathcal{R})$ theory is presented in \cite{rjim1}. The modified action for the theory is described as,	
\begin{equation}\label{action}
S=\int f(\mathcal{R},\mathcal{L}_m)	\sqrt{-g}\, d^4x,
\end{equation}	
where, $f$ represents an arbitrary function of scalar curvature $\mathcal{R}$ and the matter Lagrangian $\mathcal{L}_m$. The Energy-Momentum Tensor with an important condition, i.e., $\nabla_a \mathcal{T}^{ab}\ne0$, is obtained as a non-vanishing covariant derivative when geometry and the matter sector are explicitly coupled. As a result, test particle motion follows a non-geodesic path that affects the equivalency principle breach. Extra force orthogonal to four-velocity is caused by several forms of $\mathcal{L}_m$ representing matter sources \cite{rjim2,rjim3}. According to recent research, this idea may be taken into consideration as a potential explanation for dark energy and cosmic acceleration \cite{rjim4,rjim5}. With regard to $g^{ab}$, we can adjust Eq. (\ref{action}) to obtain the following relation for field equations.
\begin{eqnarray}
\label{fieldequation1}
f_\mathcal{R}\mathcal{R}_{ab}+(g_{ab}\nabla_a\nabla^{a}-\nabla_a\nabla_b)f_\mathcal{R}-\dfrac{1}{2}\left[f- f_{\mathcal{L}_m}\mathcal{L}_m\right]g_{ab} \nonumber\\=\dfrac{1}{2}f_{\mathcal{L}_m}\mathcal{T}_{ab}.~~
\end{eqnarray}
In the above field equations, $f_{\mathcal{L}_m}$ and $f_\mathcal{R}$ define the partial derivative of $f$ with respect to matter lagrangian $\mathcal{L}_m$ and the Ricci scalar $\mathcal{R}$ respectively. Further, $\mathcal{T}_{ab}$ define the energy-momentum tensor, which is expressed as,
\begin{equation}\label{emt}
\mathcal{T}_{ab}=-\dfrac{2}{\sqrt{-g}} \dfrac{\delta(\sqrt{-g}\mathcal{L}_m)}{\delta g^{ab}}=g_{ab}\mathcal{L}_m-2\dfrac{\partial \mathcal{L}_m}{\partial g^{ab}}.
\end{equation}
Now, by taking the covariant divergence of EMT, we can get the following expression
\begin{equation}\label{divofT}
\nabla^a \mathcal{T}_{ab}=2\left\lbrace \nabla^a \text{ln}\left[f_{\mathcal{L}_m} \right]\right\rbrace \dfrac{\partial \mathcal{L}_m }{\partial g^{ab}}.
\end{equation}
Now contracting the governing Eq. (\ref{fieldequation1}), we have the following relation between the trace of EMT and Lagrangian matter:
\begin{equation}\label{traceoffieldequation}
3\nabla_a\nabla^{a}f_\mathcal{R}+f_\mathcal{R}\mathcal{R}-2\left[f -f_{\mathcal{L}_m}\mathcal{L}_m\right]=\dfrac{1}{2}f_{\mathcal{L}_m}\mathcal{T}.
\end{equation}
Using the above equation, one can get another form of the field equation,
\begin{eqnarray}\label{fieldquation2}
f_\mathcal{R}\left( \mathcal{R}_{ab}-\dfrac{1}{3}\mathcal{R}g_{ab}\right) + \dfrac{g_{ab}}{6}\left[f -f_{\mathcal{L}_m}\mathcal{L}_m\right] \nonumber\\=\dfrac{1}{2}\left(\mathcal{T}_{ab} -\dfrac{1}{3}\mathcal{T}g_{ab}\right)f_{\mathcal{L}_m}(\mathcal{R},\mathcal{L}_m)+\nabla_a\nabla_{b}f_\mathcal{R}.
\end{eqnarray}	
The EMT for the anisotropic matter source is defined as
\begin{equation}\label{energymomentumtensor}
\mathcal{T}_{ab}=(\rho+p_\tau)\eta_a \eta_b-p_\tau\,g_{ab}+(p_r-p_\tau)\xi_{a}\xi_b,
\end{equation}
where, the 4-velocities $\eta^a$ and $\xi^a$ satisfies $\eta^a\eta_a=-1=-\xi^a\xi_a.$ The spherically symmetric space-time for the wormhole geometry is described as,
\begin{equation}\label{whmetric}
ds^2=e^{2\Phi(r)}dt^2-\dfrac{dr^2}{1-\dfrac{b(r)}{r}} - r^2\left(d\theta^2+\text{sin}^2\theta \,d\phi^2\right),
\end{equation}
where $\Phi(r)$ and $b(r)$ are defining redshift and shape functions respectively. The redshift function $\Phi$ should be finite in the entire space-time to avoid the presence of a horizon. Here in our study, to reduce the complexity of the problem, we take $\Phi$ as a constant. The radial coordinate r takes the values ranging from $r_0$ to $\infty$. The minimum value $r_0$ is called the throat radius and is the fixed point of the shape function $b(r)$, i.e., $b(r_0)=r_0$. The shape function is significant in achieving the traversability of a wormhole. It is a monotonic function and makes the space-time asymptotically flat, i.e., $\frac{b(r)}{r}$ tends to vanish for infinitely large values of the radial coordinate. Further, the shape function satisfies the flaring-out condition $\frac{b(r)-rb'(r)}{b(r)^2}>0$. At the throat, this becomes $b'(r_0)<1$. The gravitational interaction of the wormhole geometry with anisotropic matter distribution in $f(\mathcal{R},\mathcal{L}_m)$ gravity can be described by using the following field equations;
\begin{widetext}
\begin{eqnarray}
\label{fe1}(2\rho+p_r+2p_\tau)f_{\mathcal{L}_m}&&=4f_\mathcal{R}\dfrac{b'}{r^2}-(f-f_{\mathcal{L}_m}\mathcal{L}_m)\\
\label{fe2}	(-\rho-2p_r+2p_\tau)f_{\mathcal{L}_m}&&=6f_\mathcal{R}''\left(1-\dfrac{b}{r} \right)+3f_\mathcal{R}'\left(\dfrac{b-rb'}{r^2} \right)+2f_\mathcal{R}\left(\dfrac{3b-rb'}{r^3} \right)
				-(f-f_{\mathcal{L}_m}\mathcal{L}_m) \\
\label{fe3}(-\rho+p_r-p_\tau)f_{\mathcal{L}_m}&&=6\dfrac{f_\mathcal{R}''}{r}\left(1-\dfrac{b}{r} \right)-f_\mathcal{R}\left(\dfrac{3b-rb'}{r^3} \right)-(f-f_{\mathcal{L}_m}\mathcal{L}_m)
\end{eqnarray}
\end{widetext}
In this section, we shall consider a following viable $f(\mathcal{R},\mathcal{L}_m)$ model to study the characteristics of wormhole geometry:
\begin{equation}\label{mod}
f(\mathcal{R},\mathcal{L}_m)=\dfrac{\mathcal{R}}{2}+\mathcal{L}_m^\alpha
\end{equation}
where $\alpha$ is a model parameter. Now, comparing the Eqs. (\ref{fe1})-(\ref{fe3}) for $f(\mathcal{R},\mathcal{L}_m)$ model (\ref{mod}), we can get the expressions for radial and tangential pressures as
\begin{align}
\label{pr}p_r&=-\dfrac{\rho}{\alpha}\left[(\alpha-1)+\dfrac{b}{r^3\rho^\alpha}\right] \\
\label{pt}p_\tau&=\dfrac{r b'+b}{2\alpha r^3 \rho^{\alpha-1}} -\rho
\end{align}

\begin{figure*}
\centering \epsfig{file=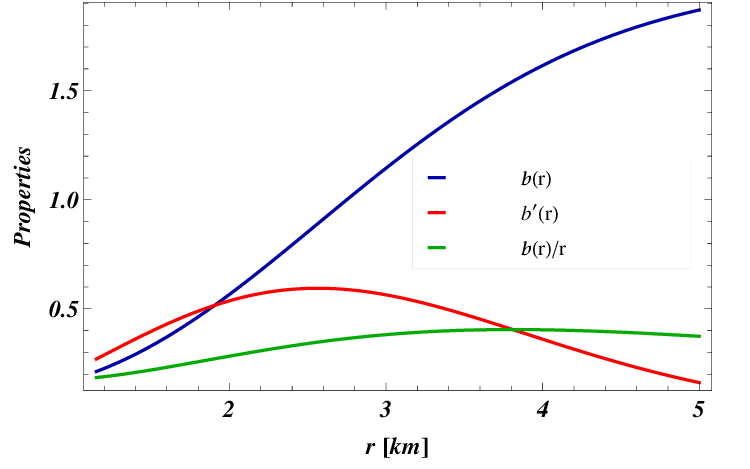, width=.45\linewidth,
height=2.8in}\epsfig{file=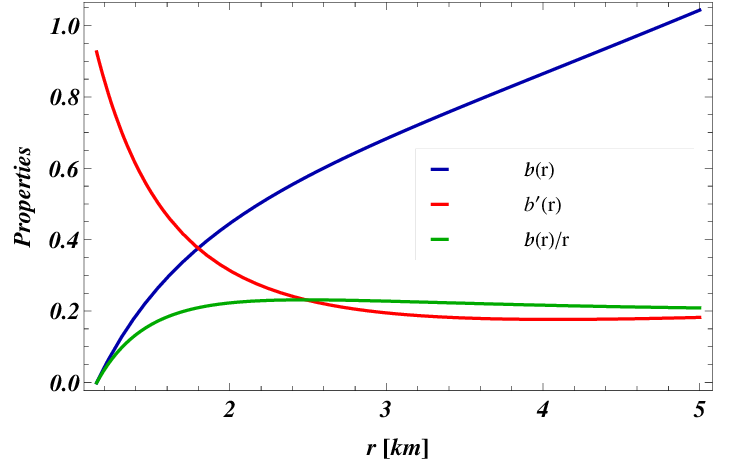, width=.45\linewidth,
height=2.8in} \caption{\label{fig1} Shows the wormhole properties for quantum wave dark matter halo (left) and CDM halo with NFW matter distribution (right). }
\end{figure*}
\section{Quantum Wave Dark Matter Halo}
A newly developed soliton matter density distribution, which is also renamed as \textit{solitonic core} in literature~\cite{DM1} and it is expressed as:
\begin{equation} \label{e1}
    \rho_\text{sc}(r) = \rho_s\left[1+\alpha\left( \frac{r}{r_s}\right)^2 \right]^{-8}.
\end{equation}
In the above equation, $\rho_s$ and $r_s$ define the soliton core density and soliton core radius respectively~\cite{DM3}. The above-said matter distribution is explored in detail in the most pioneer study~\cite{DM2}. The specific radius is predicted to be the half-density radius, which can be calculated exactly as constant $\alpha = \sqrt[8]{2}-1 \sim 0.09051$. Further, in Eq. \eqref{e1}, $\rho_s$ is provided as
\begin{equation} \label{e2}
    \rho_s = 2.4\text{x}10^{12} m_\varrho^{-2}  \left(\frac{r_s}{\text{pc}}\right)^{-4}\frac{M_\odot}{\text{pc}^{3}},
\end{equation}
where
\begin{equation}
m_\varrho = \frac{m_\text{b}}{10^{-22}\text{eV}},
\end{equation}
with $m_\text{b}$ as the boson mass parameter. The boson mass is a specific necessary parameter that can be assumed to have a single and unique value for all galaxies in the Universe~\cite{DM3}. In the current analysis, we shall derive the new wormhole solution surrounded by solitonic dark matter distribution using Eq. \eqref{e1} under the effect of the boson mass parameter. For the more generic and comparative study, we replace the exponent $8$ by $n$, i.e., in Eq. \eqref{e1} with $n$, the revised form of the solitonic core by Eq. \eqref{e1} is rewritten as:
\begin{equation} \label{e6}
\rho_\text{s}(r) = \rho_s\left[1+\alpha\left( \frac{r}{r_s}\right)^2 \right]^{-n}.
\end{equation}
We have observed that only whole-number integers can be assigned to $n$ for exploring certain cosmic scenarios. While it may be theoretically possible to choose a value of $n$ other than 8, it may not be suitable or consistent with recent observations in a physical sense. Further, for $M87$ galaxy, the values of soliton core radius is estimated $r_s = 156 \text{pc}$

\section{The CDM halo with NFW matter distribution}

The Navarro-Frenk-White (NFW) profile \cite{DM6}, which is the CDM distribution of DM halo, was discovered through analyzing $N$-body systems, which is expressed as
\begin{equation}\label{e7}
\rho(r) = \frac{\rho_s}{(r/r_s)(1+r/r_s)^2},
\end{equation}
Where $r_s$ is the typical radius and $\rho_s$ is the density at zero radius of the universe. For $M87$ galaxy, we have  $\rho_s = 0.008 \times 10^{7.5}~\text{M}_{\odot}/ \text{kpc}^3$ (see \cite{DM7}) and ${r_s} = 130~\text{kpc}$ \cite{DM7}.

\section{Wormhole study via dark matter halos}

In the current section, we shall calculate the wormhole solution within the scope of shape function by using two different kinds of dark matter halos. In this article, we plug Eq. (\ref{e6}) and Eq. (\ref{e7}) to find a new wormhole solutions that are encompassed by a distribution of solitonic dark matter and CDM halo with NFW matter distribution in the context of $f(\mathcal{R},\mathcal{L}_m)$ gravity. To accomplish this, we will directly apply the field equations of $f(\mathcal{R},\mathcal{L}_m)$ gravity and calculate the shape function while taking into account the influence of the solitonic dark matter and CDM halo with NFW matter distributions. In order to calculate the exact form of the shape function for solitonic dark matter halo, we shall plug the Eq.(\ref{e1}), Eq. (\ref{pr}), and Eq. (\ref{pt}) into Eq.(\ref{fe2}), we get the following expression for shape function
\begin{widetext}
\begin{eqnarray}\label{s1}
&& \hspace{-0.5cm}b(r) = \frac{1}{3} r^3 \left(\frac{\gamma  r^2}{r_{s}^2}+1\right)^{\alpha  n} \, _2F_1\left(\frac{3}{2},n \alpha ;\frac{5}{2};-\frac{r^2 \gamma }{r_{s}^2}\right)  \left(\rho_{s} \left(\frac{\gamma  r^2}{r_{s}^2}+1\right)^{-n}\right)^{\alpha }-\frac{1}{3} r_{0}^3 \left(\frac{\gamma  r_{0}^2}{r_{s}^2}+1\right)^{\alpha  n}  \, _2F_1\left(\frac{3}{2},n \alpha ;\frac{5}{2};-\frac{r_{0}^2 \gamma }{r_{s}^2}\right)\nonumber\\&& \hspace{0.5cm}\times \left(\rho_{s} \left(\frac{\gamma  r_{0}^2}{r_{s}^2}+1\right)^{-n}\right)^{\alpha } +r_{0},
\end{eqnarray}
\end{widetext}
where $\, _2F_1$ is representing the special function of second kind and $r_{0}$ is mentioning the wormhole throat radius. Now, the corresponding radial and pressure components are calculated as
\begin{widetext}
\begin{eqnarray}
p_{r}&&=\frac{1}{3 \alpha  r^3}\left[\left(\rho_{s} \left(\frac{\gamma  r^2}{r_{s}^2}+1\right)^{-n}\right)^{1-\alpha } \left(-r^3 \left(\frac{\gamma  r^2}{r_{s}^2}+1\right)^{\alpha  n} \, _2F_1\left(\frac{3}{2},n \alpha ;\frac{5}{2};-\frac{r^2 \gamma }{r_{s}^2}\right)\right.\right.\nonumber\\&&\left.\left.\times  \left(\rho_{s} \left(\frac{\gamma  r^2}{r_{s}^2}+1\right)^{-n}\right)^{\alpha }+r_{0}^3 \left(\frac{\gamma  r_{0}^2}{r_{s}^2}+1\right)^{\alpha  n} \, _2F_1\left(\frac{3}{2},n \alpha ;\frac{5}{2};-\frac{r_{0}^2 \gamma }{r_{s}^2}\right) \left(\rho_{s} \left(\frac{\gamma  r_{0}^2}{r_{s}^2}+1\right)^{-n}\right)^{\alpha }\right.\right.\nonumber\\&&\left.\left.-3 (\alpha -1) r^3 \left(\rho_{s} \left(\frac{\gamma  r^2}{r_{s}^2}+1\right)^{-n}\right)^{\alpha }-3 r_{0}\right)\right] ,\label{s1pr}\\
p_{t}&&=\frac{1}{6 \alpha  r^3}\left[\left(\rho_{s} \left(\frac{\gamma  r^2}{r_{s}^2}+1\right)^{-n}\right)^{1-\alpha } \left(r^3 \left(\frac{\gamma  r^2}{r_{s}^2}+1\right)^{\alpha  n} \, _2F_1\left(\frac{3}{2},n \alpha ;\frac{5}{2};-\frac{r^2 \gamma }{r_{s}^2}\right) \right.\right.\nonumber\\&&\left.\left.\left(\rho_{s} \left(\frac{\gamma  r^2}{r_{s}^2}+1\right)^{-n}\right)^{\alpha }-r_{0}^3\times \left(\frac{\gamma  r_{0}^2}{r_{s}^2}+1\right)^{\alpha  n} \, _2F_1\left(\frac{3}{2},n \alpha ;\frac{5}{2};-\frac{r_{0}^2 \gamma }{r_{s}^2}\right) \left(\rho_{s} \left(\frac{\gamma  r_{0}^2}{r_{s}^2}+1\right)^{-n}\right)^{\alpha }\right.\right.\nonumber\\&&\left.\left.-3 (2 \alpha -1) r^3 \left(\rho_{s} \left(\frac{\gamma  r^2}{r_{s}^2}+1\right)^{-n}\right)^{\alpha }+3 r_{0}\right)\right] ,\label{s1pt}
\end{eqnarray}
\end{widetext}

\begin{table*}
		\caption{Quantum Wave Dark Matter Halo with $\rho_s = 0.006 \times 10^{5.5}~\text{M}_{\odot}/ \text{kpc}^3$ and $r_s = 156 \text{pc}$}
		    \label{tab1}
		    \centering
		    \begin{tabular}{|c|c|c|}
		        \hline
		        \textit{Expressions } & \textit{Results}  & \textit{ Explanation} \\
		        \hline
		        $b(r)$ (Fig. (\ref{fig1}) left part)               &   $b(r)>0$ for $r_{0}\leq r \leq5$                 & Shape function is increasing throughout the configuration\\
		        \hline
		        $\frac{db}{dr}$(Fig. (\ref{fig1}) left part)         &   $\frac{db}{dr}\leq1$ for $r_{0}\leq r \leq5$                 & Flaring-out condition is satisfied \\
		       \hline
		       $\frac{b(r)}{r}$(Fig. (\ref{fig1}) left part)          & $ \frac{b(r)}{r}\rightarrow 0$ as $r\rightarrow \infty$                   & Asymptotic flatness condition is satisfied\\
		       \hline
         $1-\frac{b(r)}{r}$(Fig. (\ref{fig2}) left part)          & $1-\frac{b(r)}{r}\geq0$ for $r_{0}\leq r \leq5$                 & In maximum region this condition is satisfied \\
		       \hline
			\end{tabular}
		\end{table*}

\begin{table*}
		\caption{The CDM halo with NFW matter distribution with $\rho_s = 0.008 \times 10^{7.5}~\text{M}_{\odot}/ \text{kpc}^3$ and ${r_s} = 130~\text{kpc}$ }
		    \label{tab2}
		    \centering
		    \begin{tabular}{|c|c|c|}
		        \hline
		        \textit{Expressions } & \textit{Results}  & \textit{ Explanation} \\
		        \hline
		        $b(r)$ (Fig. (\ref{fig1}) right part)                 &   $b(r)>0$ for $r_{0}\leq r \leq5$                 & Shape function is increasing throughout the configuration\\
		        \hline
		        $\frac{db}{dr}$ (Fig. (\ref{fig1}) right part)        &   $\frac{db}{dr}\leq1$ for $r_{0}\leq r \leq5$                 & Flaring-out condition is satisfied \\
		       \hline
		       $\frac{b(r)}{r}$ (Fig. (\ref{fig1}) right part)         & $ \frac{b(r)}{r}\rightarrow 0$ as $r\rightarrow \infty$                   & Asymptotic flatness condition is satisfied\\
		       \hline
          $1-\frac{b(r)}{r}$(Fig. (\ref{fig2}) right part)          & $1-\frac{b(r)}{r}\geq0$ for $r_{0}\leq r \leq5$                 & In maximum region this condition is satisfied \\
		       \hline
			\end{tabular}
		\end{table*}
  
\begin{figure*}
\centering \epsfig{file=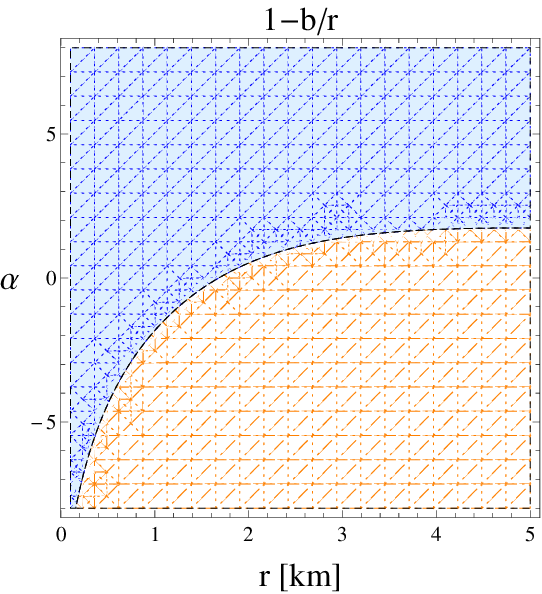, width=.45\linewidth,
height=2.2in}~~~~~~\epsfig{file=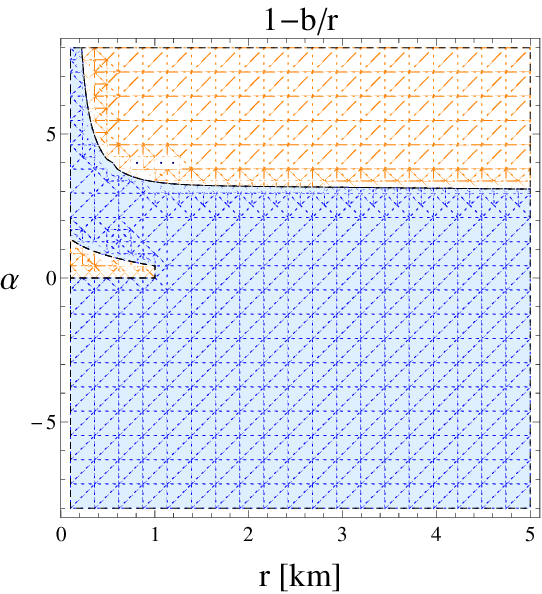, width=.45\linewidth,
height=2.2in} \caption{\label{fig2} Shows the valid region${}^{(\textcolor{blue}{\bigstar})}$ and invalid region${}^{(\textcolor{orange}{\bigstar})}$ for both calculated wormhole solutions with quantum wave dark matter halo (left) and CDM halo with NFW matter distribution (right).}
\end{figure*}
\begin{figure*}
\centering \epsfig{file=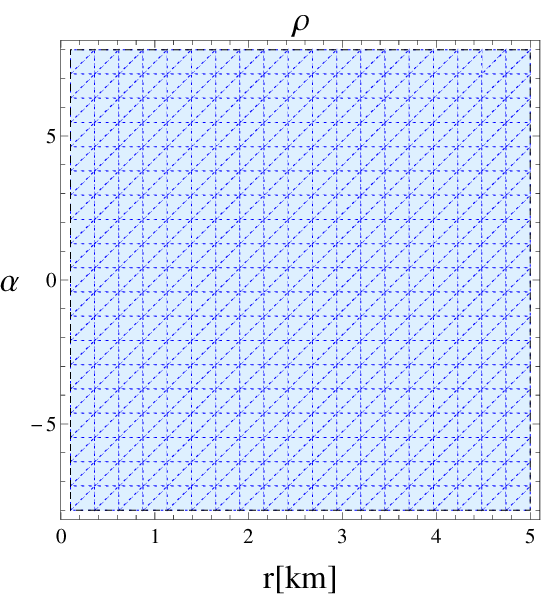, width=.45\linewidth,
height=2.2in}~~~~~~\epsfig{file=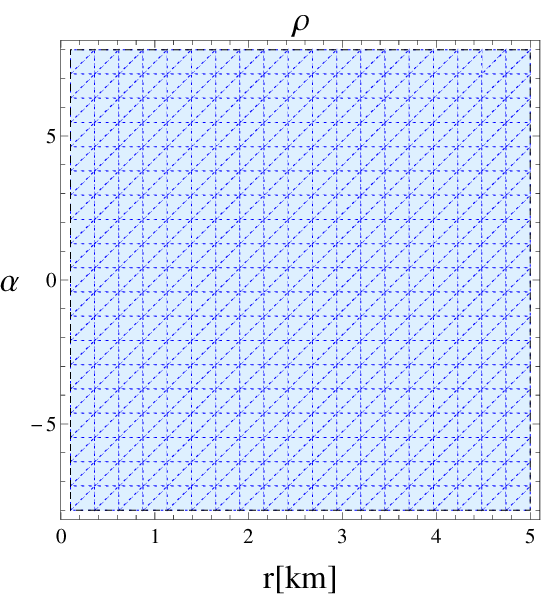, width=.45\linewidth,
height=2.2in} \caption{\label{fig3} Shows the valid region${}^{(\textcolor{blue}{\bigstar})}$ and invalid region${}^{(\textcolor{orange}{\bigstar})}$ for energy density against both calculated wormhole solutions with quantum wave dark matter halo (left) and CDM halo with NFW matter distribution (right).}
\end{figure*}
\begin{figure*}
\centering \epsfig{file=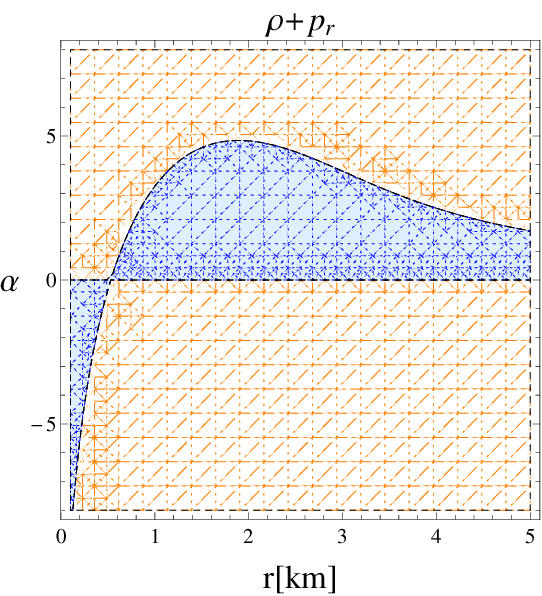, width=.45\linewidth,
height=2.2in}~~~~~~\epsfig{file=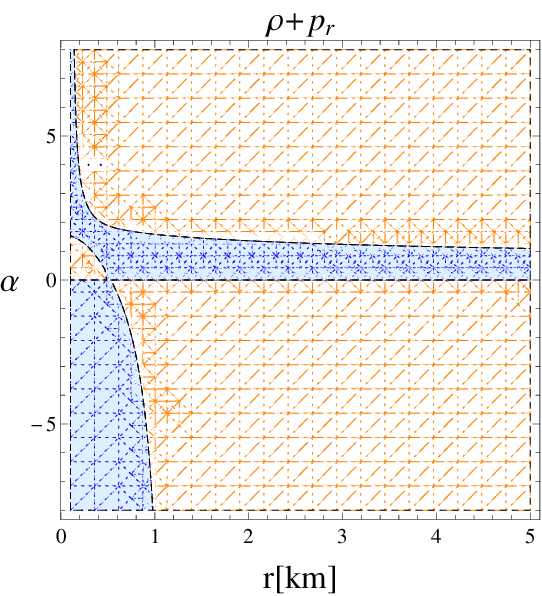, width=.45\linewidth,
height=2.2in} \caption{\label{fig4} Shows the valid region${}^{(\textcolor{blue}{\bigstar})}$ and invalid region${}^{(\textcolor{orange}{\bigstar})}$ for $\rho+p_r$ against both calculated wormhole solutions with quantum wave dark matter halo (left) and CDM halo with NFW matter distribution (right).}
\end{figure*}
\begin{figure*}
\centering \epsfig{file=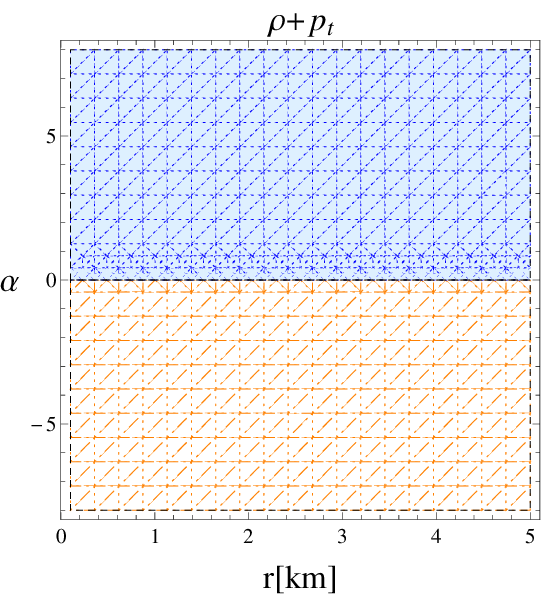, width=.45\linewidth,
height=2.2in}~~~~~~\epsfig{file=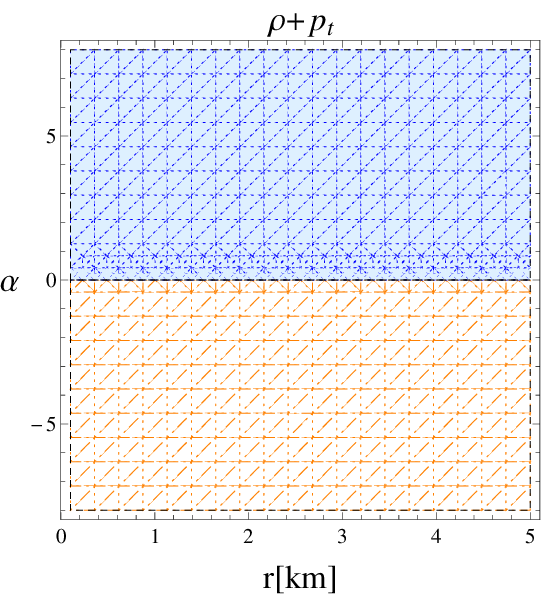, width=.45\linewidth,
height=2.2in} \caption{\label{fig5} Shows the valid region${}^{(\textcolor{blue}{\bigstar})}$ and invalid region${}^{(\textcolor{orange}{\bigstar})}$ for $\rho+p_t$ against both calculated wormhole solutions with quantum wave dark matter halo (left) and CDM halo with NFW matter distribution (right).}
\end{figure*}

\begin{table*}
		\caption{Quantum Wave Dark Matter Halo with $\rho_s = 0.006 \times 10^{5.5}~\text{M}_{\odot}/ \text{kpc}^3$ and $r_s = 156 \text{pc}$}
		    \label{tab3}
		    \centering
		    \begin{tabular}{|c|c|c|}
		        \hline
		        \textit{Expressions } & \textit{Results}  & \textit{ Explanation} \\
		        \hline
		        $\rho$ (Fig. (\ref{fig3}) left part)               &   $\rho>0$ for $r_{0}\leq r \leq5$                 & Energy density is observed positive\\
		        \hline
		        $\rho+p_r$(Fig. (\ref{fig4}) left part)         &  $\rho+p_r <0$ for $r_{0}\leq r \leq5$                 & Maximum supportive region${}^{(\textcolor{orange}{\bigstar})}$ for wormhole existence \\
		       \hline
		       $\rho+p_t$(Fig. (\ref{fig5}) left part)          & $\rho+p_t<0$ for $r_{0}\leq r \leq5$                   & Maximum supportive region${}^{(\textcolor{orange}{\bigstar})}$ for wormhole existence \\
          \hline
         $\rho-p_r$(Fig. (\ref{fig6}) left part)          &  $\rho-p_r<0$ for $r_{0}\leq r \leq5$                 & Minimum supportive region${}^{(\textcolor{orange}{\bigstar})}$ for wormhole existence\\
		      		       \hline
                  $\rho-p_t$(Fig. (\ref{fig7}) left part)          & $\rho-p_t<0$ for $r_{0}\leq r \leq5$                  & Minimum supportive region${}^{(\textcolor{orange}{\bigstar})}$ for wormhole existence\\
		      
		       \hline
          $\rho-|p_r|$(Fig. (\ref{fig8}) left part)          & $\rho-|p_r|<0$ for $r_{0}\leq r \leq5$                  & Maximum supportive region${}^{(\textcolor{orange}{\bigstar})}$ for wormhole existence\\
		      
		       \hline
         $\rho-|p_t|$(Fig. (\ref{fig9}) left part)          & $\rho-|p_t|<0$ for $r_{0}\leq r \leq5$                   & Maximum supportive region${}^{(\textcolor{orange}{\bigstar})}$ for wormhole existence\\
		      
		       \hline
          $\rho+p_{r}+2p_{t}$(Fig. (\ref{fig10}) left part)          &  $\rho+p_{r}+2p_{t}<0$ for $r_{0}\leq r \leq5$              & Maximum supportive region${}^{(\textcolor{orange}{\bigstar})}$ for wormhole existence\\
		      
		       \hline
			\end{tabular}
		\end{table*}

\begin{table*}
		\caption{The CDM halo with NFW matter distribution with $\rho_s = 0.008 \times 10^{7.5}~\text{M}_{\odot}/ \text{kpc}^3$ and ${r_s} = 130~\text{kpc}$ }
		    \label{tab4}
		    \centering
		    \begin{tabular}{|c|c|c|}
		       		         \hline
		        \textit{Expressions } & \textit{Results}  & \textit{ Explanation} \\
		        \hline
		        $\rho$ (Fig. (\ref{fig3}) right part)               &   $\rho>0$ for $r_{0}\leq r \leq5$                 & Energy density is observed positive\\
		        \hline
		        $\rho+p_r$(Fig. (\ref{fig4}) right part)         &  $\rho+p_r <0$ for $r_{0}\leq r \leq5$                 & Maximum supportive region${}^{(\textcolor{orange}{\bigstar})}$ for wormhole existence \\
		       \hline
		       $\rho+p_t$(Fig. (\ref{fig5}) right part)          & $\rho+p_t<0$                  & Maximum supportive region${}^{(\textcolor{orange}{\bigstar})}$ for wormhole existence \\
          \hline
         $\rho-p_r$(Fig. (\ref{fig6}) right part)          &  $\rho-p_r<0$ for $r_{0}\leq r \leq5$                 & Minimum supportive region${}^{(\textcolor{orange}{\bigstar})}$ for wormhole existence\\
		      		       \hline
                  $\rho-p_t$(Fig. (\ref{fig7}) right part)          & $\rho-p_t<0$ for $r_{0}\leq r \leq5$                  & Minimum supportive region${}^{(\textcolor{orange}{\bigstar})}$ for wormhole existence\\
		      
		       \hline
          $\rho-|p_r|$(Fig. (\ref{fig8}) right part)          & $\rho-|p_r|<0$ for $r_{0}\leq r \leq5$                  & Maximum supportive region${}^{(\textcolor{orange}{\bigstar})}$ for wormhole existence\\
		      
		       \hline
         $\rho-|p_t|$(Fig. (\ref{fig9}) right part)          & $\rho-|p_t|<0$ for $r_{0}\leq r \leq5$                   & Maximum supportive region${}^{(\textcolor{orange}{\bigstar})}$ for wormhole existence\\
		      
		       \hline
          $\rho+p_{r}+2p_{t}$(Fig. (\ref{fig10}) right part)          &  $\rho+p_{r}+2p_{t}<0$ for $r_{0}\leq r \leq5$              & Maximum supportive region${}^{(\textcolor{orange}{\bigstar})}$ for wormhole existence\\
		      
		       \hline
			\end{tabular}
		\end{table*}

\begin{figure*}
\centering \epsfig{file=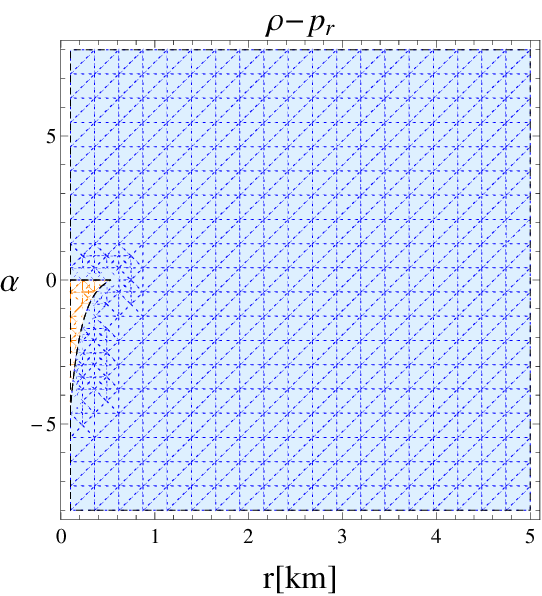, width=.45\linewidth,
height=2.2in}~~~~~~\epsfig{file=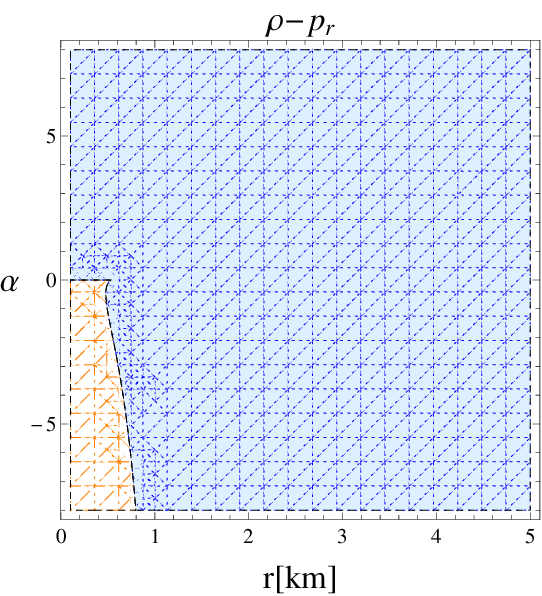, width=.45\linewidth,
height=2.2in} \caption{\label{fig6} Shows the valid region${}^{(\textcolor{blue}{\bigstar})}$ and invalid region${}^{(\textcolor{orange}{\bigstar})}$ for $\rho-p_r$ against both calculated wormhole solutions with quantum wave dark matter halo (left) and CDM halo with NFW matter distribution (right).}
\end{figure*}
\begin{figure*}
\centering \epsfig{file=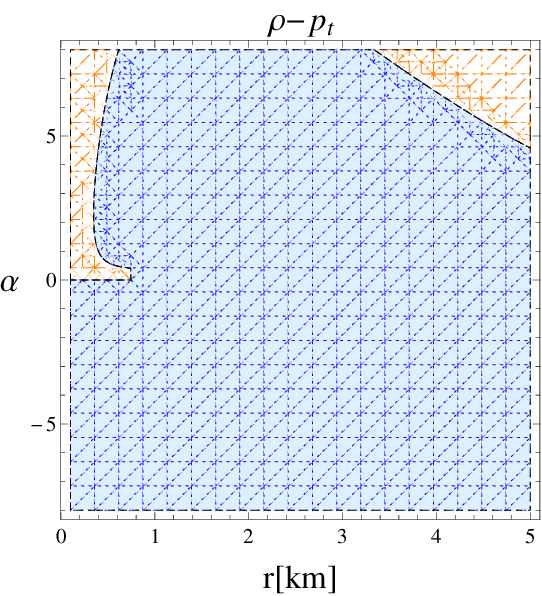, width=.45\linewidth,
height=2.2in}~~~~~~\epsfig{file=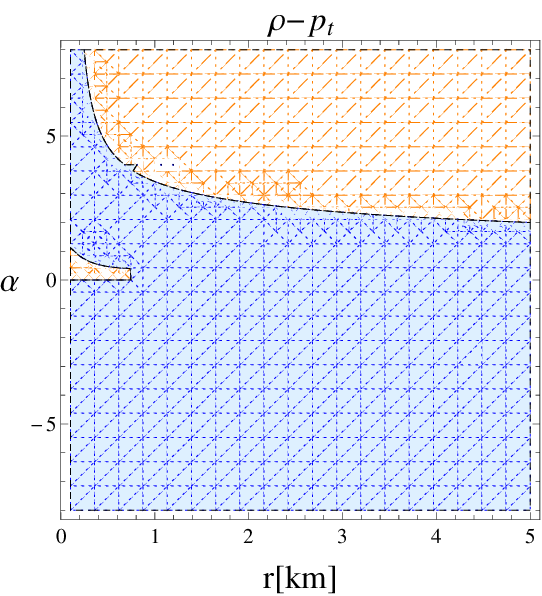, width=.45\linewidth,
height=2.2in} \caption{\label{fig7} Shows the valid region${}^{(\textcolor{blue}{\bigstar})}$ and invalid region${}^{(\textcolor{orange}{\bigstar})}$ for $\rho-p_t$ against both calculated wormhole solutions with quantum wave dark matter halo (left) and CDM halo with NFW matter distribution (right).}
\end{figure*}
\begin{figure*}
\centering \epsfig{file=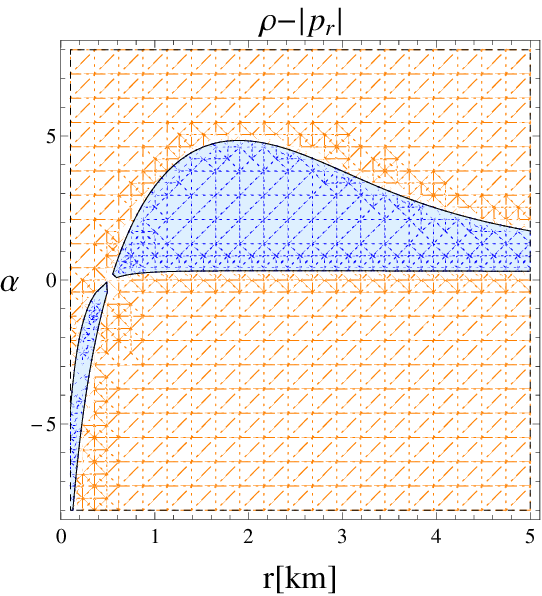, width=.45\linewidth,
height=2.2in}~~~~~~\epsfig{file=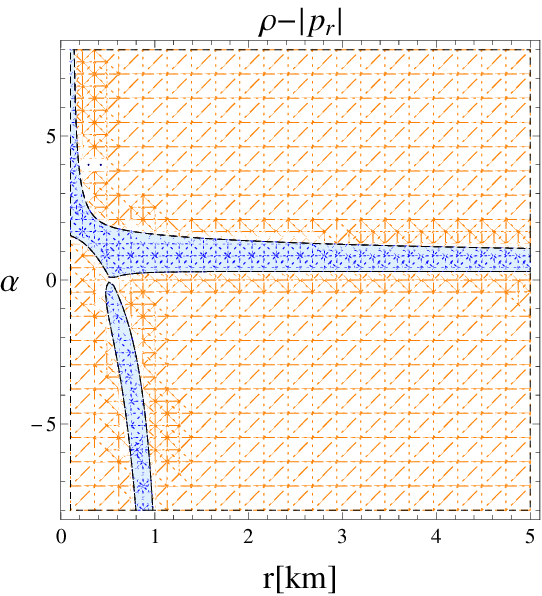, width=.45\linewidth,
height=2.2in} \caption{\label{fig8} Shows the valid region${}^{(\textcolor{blue}{\bigstar})}$ and invalid region${}^{(\textcolor{orange}{\bigstar})}$ for $\rho+|p_r|$ against both calculated wormhole solutions with quantum wave dark matter halo (left) and CDM halo with NFW matter distribution (right).}
\end{figure*}
\begin{figure*}
\centering \epsfig{file=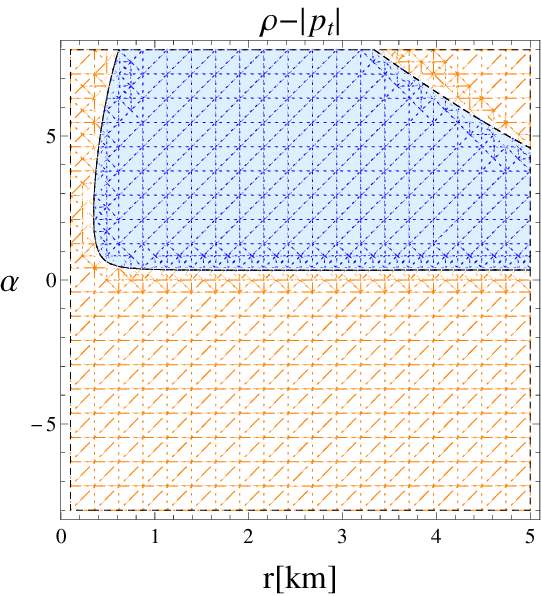, width=.45\linewidth,
height=2.2in}~~~~~~\epsfig{file=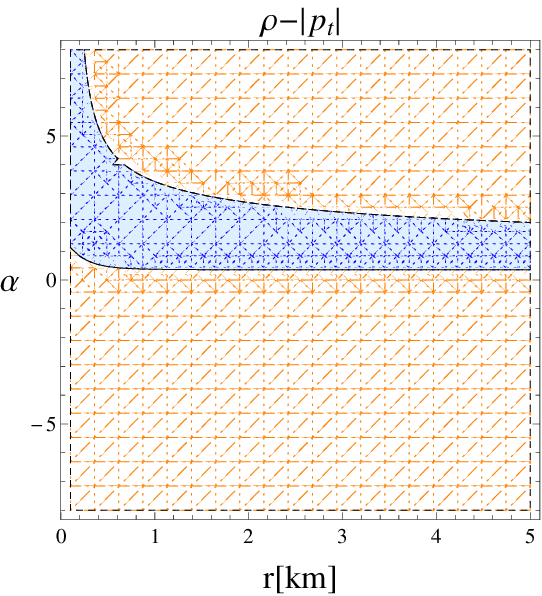, width=.45\linewidth,
height=2.2in} \caption{\label{fig9}Shows the valid region${}^{(\textcolor{blue}{\bigstar})}$ and invalid region${}^{(\textcolor{orange}{\bigstar})}$ for $\rho+|p_t|$ against both calculated wormhole solutions with quantum wave dark matter halo (left) and CDM halo with NFW matter distribution (right).}
\end{figure*}
\begin{figure*}
\centering \epsfig{file=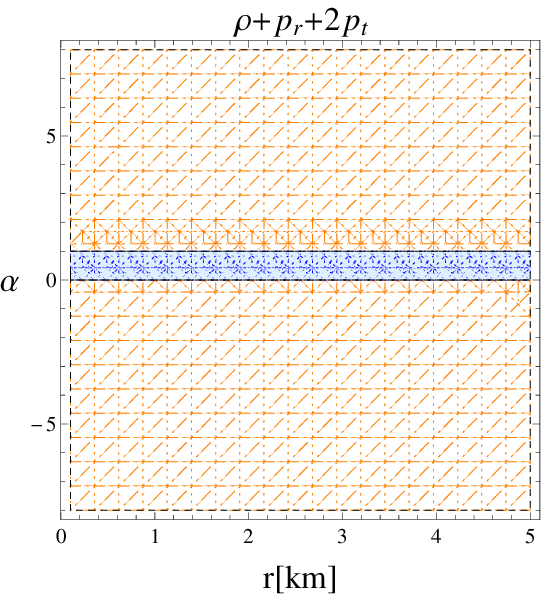, width=.45\linewidth,
height=2.2in}~~~~~~\epsfig{file=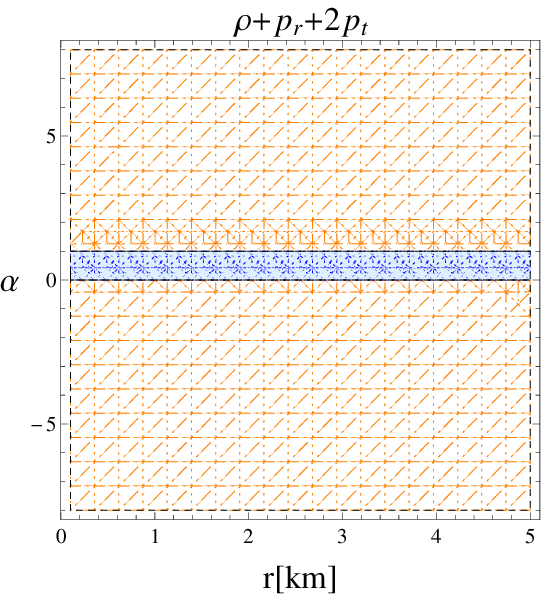, width=.45\linewidth,
height=2.2in} \caption{\label{fig10} Shows the valid region${}^{(\textcolor{blue}{\bigstar})}$ and invalid region${}^{(\textcolor{orange}{\bigstar})}$ for $\rho+p_r+2p_{t}$ against both calculated wormhole solutions with quantum wave dark matter halo (left) and CDM halo with NFW matter distribution (right).}
\end{figure*}

\begin{figure*}
\centering
\epsfig{file=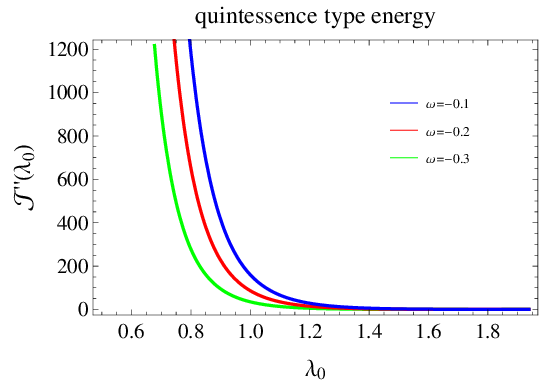,width=.45\linewidth}\epsfig{file=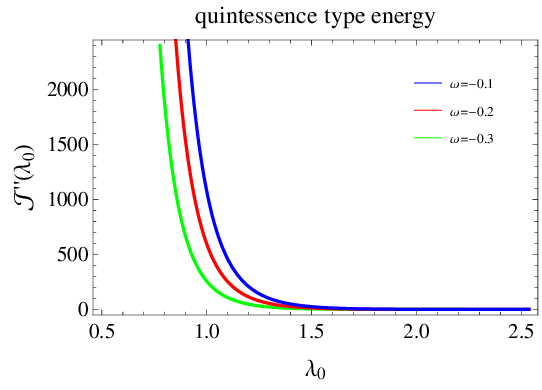,width=.45\linewidth}
\caption{\label{f1} Dynamical stability of the shell for the case of CDM halo matter composed with quintessence energy for different values of
$\omega$ and $r_0=0.5$ (left plot), $r_0=1.5$ (right plot) with
$\alpha =-0.9, \gamma =0.09051,  m=0.5 n=8$.}
\epsfig{file=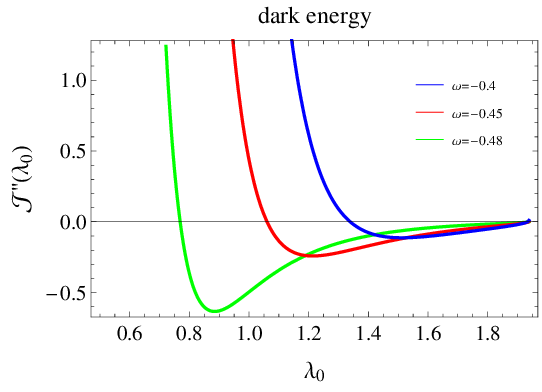,width=.45\linewidth}\epsfig{file=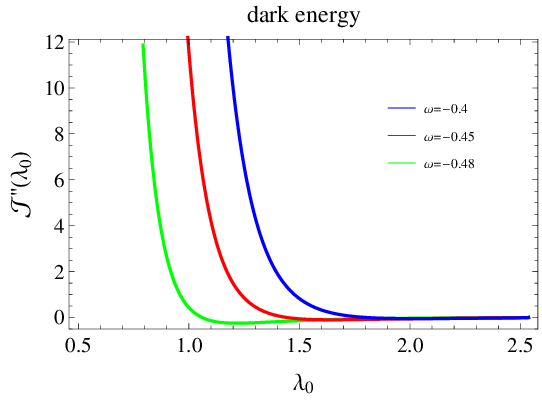,width=.45\linewidth}
\caption{\label{f2} Dynamical stability of the shell for the case of CDM halo matter composed with dark energy for different values of $\omega$
and $r_0=0.5$ (left plot), $r_0=1.5$ (right plot) with $\alpha =-0.9, \gamma =0.09051,  m=0.5 n=8$.}
\epsfig{file=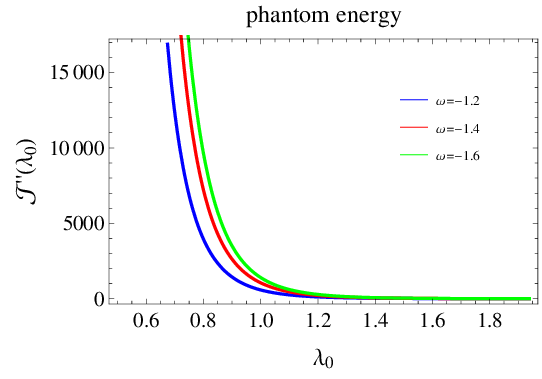,width=.45\linewidth}\epsfig{file=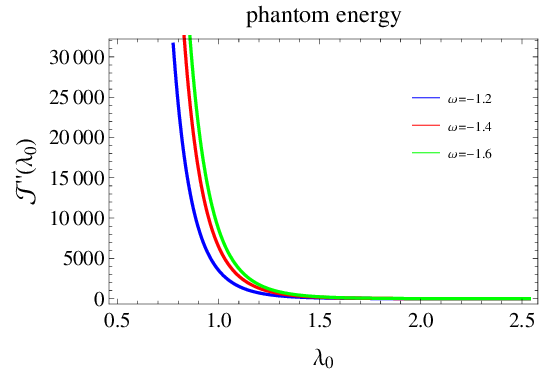,width=.45\linewidth}
\caption{\label{f3} Dynamical stability of the shell for the case of CDM halo matter composed with phantom energy for different values of
$\omega$ and $r_0=0.5$ (left plot), $r_0=1.5$ (right plot) with
$\alpha =-0.9, \gamma =0.09051,  m=0.5 n=8$.}
\end{figure*}

\begin{figure*}\centering
\epsfig{file=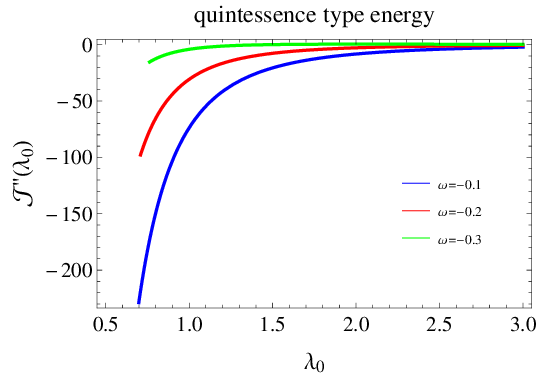,width=.45\linewidth}\epsfig{file=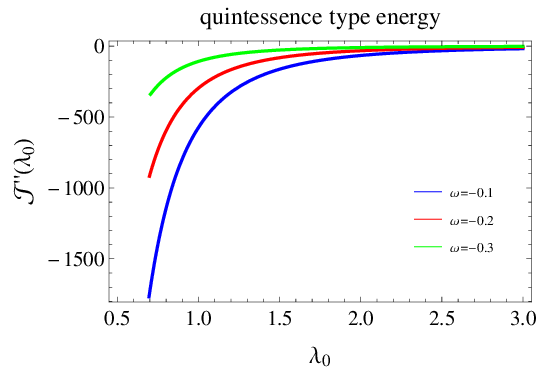,width=.45\linewidth}
\caption{\label{f4} Dynamical stability of the shell for the case of quantum wae dark matter composed with quintessence energy for different values of
$\omega$ and $r_0=0.5$ (left plot), $r_0=1.5$ (right plot) with
$\alpha =-0.9, \gamma =0.09051,  m=0.5 n=8$..}
\epsfig{file=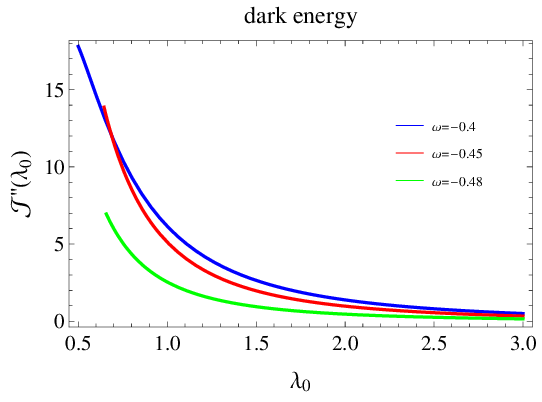,width=.45\linewidth}\epsfig{file=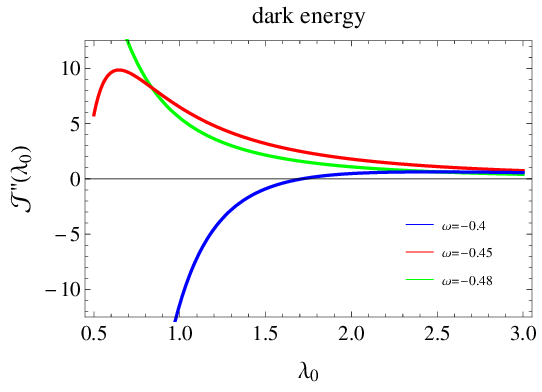,width=.45\linewidth}
\caption{\label{f5} Dynamical stability of the shell for the case of quantum wae dark matter composed with dark energy for different values of $\omega$
and $r_0=0.5$ (left plot), $r_0=1.5$ (right plot) with $\alpha =-0.9, \gamma =0.09051,  m=0.5 n=8$.}
\epsfig{file=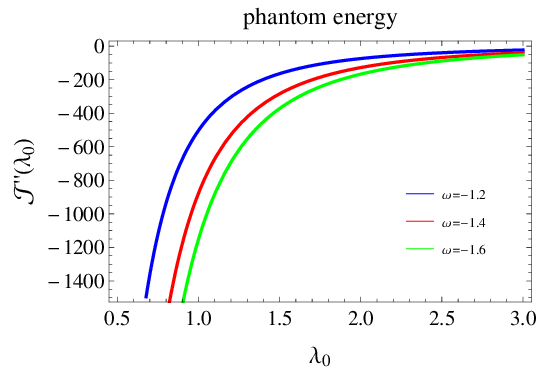,width=.45\linewidth}\epsfig{file=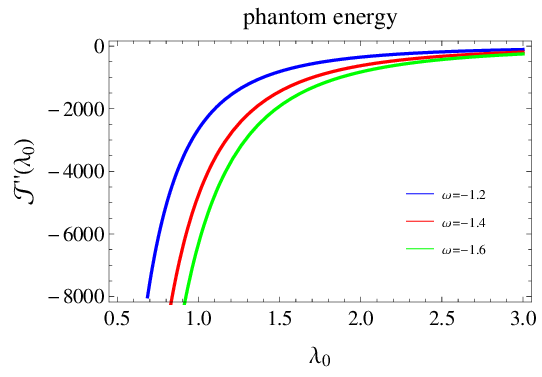,width=.45\linewidth}
\caption{\label{f6} Dynamical stability of the shell for the case of quantum wae dark matter composed with phantom energy for different values of $\omega$ and $r_0=0.5$ (left plot), $r_0=1.5$ (right plot) with
$\alpha =-0.9, \gamma =0.09051,  m=0.5 n=8$.}
\end{figure*}
Now, again by using Eq.(\ref{e2}), Eq. (\ref{pr}), and Eq. (\ref{pt}) into Eq.(\ref{fe2}), one can get the exact shape function for CDM halo with NFW matter distribution, which is calculated as
\begin{widetext}
\begin{eqnarray}\label{s2}
b(r)&&= \frac{1}{2 \alpha  r^4 (r_{s}+r)^2}\bigg(\rho_{s} r_{s}^2 \left(\Gamma (3-\alpha ) \left(r^3 (r_{s}+r) \, _2\tilde{F}_1\left(1,4-3 \alpha ;4-\alpha ;-\frac{r}{r_{s}}\right)-r_{0}^3 (r_{s}+r_{0}) \right.\right.\nonumber\\&&\times\left.\left. \, _2\tilde{F}_1\left(1,4-3 \alpha ;4-\alpha ;-\frac{r_{0}}{r_{s}}\right) \left(\frac{\rho_{s} r_{s}^3}{r (r_{s}+r)^2}\right)^{-\alpha } \left(\frac{\rho_{s} r_{s}^3}{r_{0} (r_{s}+r_{0})^2}\right)^{\alpha }\right)+r_{s} \left(r_{0} \left(\frac{\rho_{s} r_{s}^3}{r (r_{s}+r)^2}\right)^{-\alpha }\right.\right.\nonumber\\&&\left.\left.+(1-2 \alpha ) r^3\right)\right)\bigg),
\end{eqnarray}
\end{widetext}
where $\, _2F_1$ is representing the special function of second kind and $r_{0}$ is mentioning the wormhole throat radius. Now, the corresponding radial and pressure components are calculated as
\begin{widetext}
\begin{eqnarray}
p_{r}&&=\frac{1}{\alpha  r^4 (r_{s}+r)^2}\left[\rho_{s} r_{s}^2 \left(\Gamma (3-\alpha ) \left(r_{0}^3 (r_{s}+r_{0}) \, _2\tilde{F}_1\left(1,4-3 \alpha ;4-\alpha ;-\frac{r_{0}}{r_{s}}\right) \left(\frac{\rho_{s} r_{s}^3}{r (r_{s}+r)^2}\right)^{-\alpha } \right.\right.\right.\nonumber\\&&\times\left.\left.\left.\left(\frac{\rho_{s} r_{s}^3}{r_{0} (r_{s}+r_{0})^2}\right)^{\alpha }-r^3 (r_{s}+r) \, _2\tilde{F}_1\left(1,4-3 \alpha ;4-\alpha ;-\frac{r}{r_{s}}\right)\right)+r_{s} \left(r_{0} \left(-\left(\frac{\rho_{s} r_{s}^3}{r (r_{s}+r)^2}\right)^{-\alpha }\right)\right.\right.\right.\nonumber\\&&-\left.\left.\left.(\alpha -1) r^3\right)\right)\right] ,\label{s2pr}\\
p_{t}&&=\frac{1}{2 \alpha  r^4 (r_{s}+r)^2}\left[\rho_{s} r_{s}^2 \left(\Gamma (3-\alpha ) \left(r^3 (r_{s}+r) \, _2\tilde{F}_1\left(1,4-3 \alpha ;4-\alpha ;-\frac{r}{r_{s}}\right)-r_{0}^3 (r_{s}+r_{0})\right.\right.\right.\nonumber\\&&\times\left.\left.\left. \, _2\tilde{F}_1\left(1,4-3 \alpha ;4-\alpha ;-\frac{r_{0}}{r_{s}}\right) \left(\frac{\rho_{s} r_{s}^3}{r (r_{s}+r)^2}\right)^{-\alpha } \left(\frac{\rho_{s} r_{s}^3}{r_{0} (r_{s}+r_{0})^2}\right)^{\alpha }\right)+r_{s} \left(r_{0} \left(\frac{\rho_{s} r_{s}^3}{r (r_{s}+r)^2}\right)^{-\alpha }\right.\right.\right.\nonumber\\&&+\left.\left.\left.(1-2 \alpha ) r^3\right)\right)\right] ,\label{s2pt}
\end{eqnarray}
\end{widetext}
Now, we are going to discuss the behavior of calculated shape functions and valid regions for the existence of wormhole solutions under the effect of two different dark matter halos. Both the calculated shape functions through Eq. (\ref{s1}) and Eq. (\ref{s2}) satisfy the required criteria for the existence of wormhole geometry. From Fig. (\ref{fig1}), one can confirm the graphical behavior for wormhole properties under the quantum wave dark matter halo effect in the left penal and for CDM halo with NFW matter distribution in the right penal. In both parts of Fig. (\ref{fig1}), the calculated shape functions are increasing and remain positive throughout the configurations. The flaring out condition, i.e., derivative of the calculated shape functions with respect to radial coordinate $r$, is also satisfied under the effect of both considered dark matter halos effect. It is also observed from the Fig. (\ref{fig1}) that both the calculated wormhole solutions are asymptotically flat under the effect of quantum wave dark matter halo and CDM halo with NFW matter distributions. Further, the valid region for both calculated wormhole solutions is also presented in Fig. (\ref{fig2}) with blue shaded against the specific range of model parameter $\alpha$. 
\section{Matter Distribution and Energy Conditions}

In this section, we shall deal with the matter distributions through energy conditions in the background of $f(\mathcal{R},\mathcal{L}_m)$ gravity. The energy conditions let us closely examine the ordinary and geodesic structure of space-time, making them essential tools for $f(\mathcal{R},\mathcal{L}_m)$ gravity. The primary energy limitations are the dominant energy condition $(DEC)$, strong energy condition $(SEC)$, weak energy condition $(WEC)$, and null energy condition $(NEC)$. 
In relation to main pressure, the ensuing energy conditions are defined as SEC $\Leftrightarrow \forall j,\rho+p_j\geq 0,\rho+\sum_jp_j\geq 0, $, DEC $\Leftrightarrow\rho\geq 0,\forall j,p_j\epsilon[-\rho,+\rho]$,
NEC$\Leftrightarrow\forall j,\rho+p_j\geq 0,$ and WEC$\Leftrightarrow\rho\geq 0,\forall j,\rho+p_j\geq 0.$ The final version of the energy conditions with regard to principal pressures are expressed as SEC $\rho+p_{r}\geq 0,~\rho+p_t\geq 0,~\rho+p_r+2p_t\geq 0$,  DEC $\rho\geq 0,~\rho-|p_r|\geq 0,~\rho-|p_t|\geq 0$,  NEC $\rho+p_r\geq 0,~\rho+p_t\geq 0$, and WEC $\rho\geq 0,~\rho+p_r\geq 0,~\rho+p_t\geq 0.$\\

The above-mentioned energy bounds have an important role in the existence of wormhole solutions. To understand the nature of involved matter, all the above energy bounds leads us to confirm the presence of exotic matter which is a necessary component for the throat radius pf wormhole geometry. Overall, research into wormhole geometry and associated energy conditions is critical for understanding the theoretical and physical consequences of these fascinating space-time structures, as well as establishing if wormholes may exist in the universe. The current analysis shows all the energy conditions showing their valid and invalid regions for the different ranges of involved model parameter $\alpha$. The energy density with $-8\leq\alpha\leq8$ remains positive under the effect of two different kinds of dark matter halos, i.e., quantum wave dark matter halo and CDM halo with NFW matter distribution. The valid region for energy density can be seen in Fig (\ref{fig3}). The NEC, i.e., $\rho+p_r$, is strongly violated for $-8\leq\alpha\leq8$ due to the maximum invalid region, which can be confirmed from Fig. (\ref{fig4}). The NEC violation is a positive indication of the presence of exotic matter. The presence of exotic matter allows the wormhole to exist. All the other energy conditions, including SEC, are presented in Figs. (\ref{fig5}-\ref{fig10}) for $-8\leq\alpha\leq8$. It is noticed from the Figs. (\ref{fig5}-\ref{fig10}) that all the energy conditions are violated in maximum regions for $-8\leq\alpha\leq8$. The invalid region for SEC from Fig. (\ref{fig10}) is another strong indication of exotic matter's presence, which is a necessary part of wormhole physics. In the context of wormholes, violating the energy conditions might result in the formation of exotic matter with negative energy conditions, which are required to keep the wormhole stable and from collapsing.

\section{Impact of quantum wave dark matter and CDM halo matter distribution on the stability of the shell}

Here, we are interested in exploring the impact of quantum wave dark matter and CDM halo matter distribution on the stability of the shell around the obtained wormhole solutions. It is interesting to mention that the supermassive black holes' gravitational ringings can be marked by the presence of dark matter. Dark matter encircles the black hole in a spike configuration, altering the local space-time and causing gravity waves released during ringdown to have unique characteristics \cite{th1}. When dark matter spikes are present, the isospectrality, which describes how black holes' quasi-normal modes are identical, is violated \cite{th2,th3}. Dark matter black holes and holes without dark matter can have very different ringing frequencies, with the relative difference reaching as high as $10^2$ \cite{th4}. These characteristics can be used for gravitational wave detection in the future to investigate if supermassive black holes have any surrounding dark matter. Gravitational wave emission may also identify a dense dark matter clump that forms when dark matter particles enter the ergosphere of a supermassive black hole. Black holes and dark matter are explored in quantum wave dark matter. The effect of fuzzy dark matter (FDM) on supermassive black holes (SMBH) is investigated, considering a soliton core density profile surrounding the SMBH \cite{fg1}. The behavior of the shadow radius and the thin-accretion disk are analyzed, showing deviations and varying luminosity near the event horizon \cite{fg1}, \cite{fg2}. Compact stellar-size particles' inspiral into these dark matter structures can yield helpful information regarding dark matter's nature. 

We consider two black hole solutions for the quantum wave dark matter and CDM halo matter distribution. Then, we match these manifolds with the respective obtained wormhole solutions. For this purpose, we assume wormhole geometry as an interior manifold and black hole solution as an outer manifold. By taking the black hole as an outer space-time, the line element of the outer manifold (+) is given as
\begin{equation}\label{1kk}
ds^{2}_+=-\mathcal{F}_+(r_+)^{-1}dr^{2}_+-r^{2}_+
d\theta^{2}_+-r^{2}_+\sin^2{\theta}_+
d\phi^{2}_++\mathcal{F}_+(r_+)dt^{2}_+.
\end{equation}
Also, the inner manifold (-) in the form of the wormhole is given as  
\begin{equation}\label{3kk}
ds^2_-=-e^{2\phi(r_-)}dt^2_-+r^2_-d\theta^2_-+\frac{dr^2_-}{\mathcal{G}_-(r_-)}+r^2_-sin^2\theta_-
d\phi^2_-.
\end{equation}
In order to develop a thin-shell around the WH geometry, we consider
two cases as follows:
\begin{itemize}

\item  For CDM halo matter distribution, we consider the lapse function of CDM black hole and the respective wormhole temporal coordinate are considered as \cite{th5}
\begin{equation}\label{2kk1}
\mathcal{F}_{+}(r_+)=\left(\frac{r_{+}}{r_{s}}+1\right)^{-\frac{8 \pi  \rho_s
r_{s}^3}{r}}-\frac{2 m}{r_{+}},
\end{equation}
 and
\begin{widetext}
\begin{eqnarray}\nonumber
\mathcal{G}_{-}(r_-)&=&1-\frac{\rho_s r_{s}^2 }{2 \alpha  r^5_{-}
(r_{s}+r_{-})^2}\left(\Gamma (3-\alpha ) \left(r^3_{-} (r_{s}+r_{-}) \,
_2\tilde{F}_1\left(1,4-3 \alpha ;4-\alpha
;-\frac{r_{-}}{r_{s}}\right)-r_{0}^3 (r_{s}+r_{0})
\right.\right.\\\nonumber&\times&\left.\left. \,
_2\tilde{F}_1\left(1,4-3 \alpha ;4-\alpha
;-\frac{r_{0}}{r_{s}}\right) \left(\frac{\rho_s r_{s}^3}{r_{-}
(r_{s}+r_{-})^2}\right)^{-\alpha } \left(\frac{\rho_s
r_{s}^3}{r_{0} (r_{s}+r_{0})^2}\right)^{\alpha
}\right)\right.\\\label{4kk}&+&\left.r_{s} \left(r_{0}
\left(\frac{\rho_s r_{s}^3}{r (r_{s}+r_{-})^2}\right)^{-\alpha }+(1-2
\alpha ) r^3_{-}\right)\right),
\end{eqnarray}
\end{widetext}
respectively. Here, $m$ is the mass of the black hole.

\item  For quantum wave dark matter, we consider the lapse functions of a black hole in quantum wave dark matter, and the respective wormhole temporal coordinates are considered as \cite{fg1}
\begin{equation}\label{2kk}
\mathcal{F}_{+}(r_+)=-\frac{4 \pi  \rho_s r_{s}^3}{7 \alpha
r_{s}}-\frac{2 m}{r_{+}}+1,
\end{equation}
and
\begin{widetext}
\begin{eqnarray}\nonumber
\mathcal{G}_{-}(r_-)&=&1-\frac{1}{3 r_{-}}\left(r^3 _{-}\left(\frac{\gamma
r^2_{-}}{r_{s}^2}+1\right)^{\alpha  n} \, _2F_1\left(\frac{3}{2},n
\alpha ;\frac{5}{2};-\frac{r^2_{-} \gamma }{r_{s}^2}\right) \left(\rho_s
\left(\frac{\gamma  r^2_{-}}{r_{s}^2}+1\right)^{-n}\right)^{\alpha
}-r_{0}^3 \left(\frac{\gamma
r_{0}^2}{r_{s}^2}+1\right)^{\alpha  n}
\right.\\\label{4kk1}&\times&\left.\, _2F_1\left(\frac{3}{2},n \alpha
;\frac{5}{2};-\frac{r_{0}^2 \gamma }{r_{s}^2}\right) \left(\rho_s
\left(\frac{\gamma
r_{0}^2}{r_{s}^2}+1\right)^{-n}\right)^{\alpha }+3
r_{0}\right),
\end{eqnarray}
\end{widetext}
respectively. 

\end{itemize}

The created framework is protected from singularities and event horizons by connecting these space-times using a well-established cut-and-paste method. A thin-shell space-time enclosing a wormhole can modify its geometry using this approach. This leads to the following divisions of the manifold: Mathematical matrix $\mathcal{M}^{\pm}=\left\lbrace r^{\pm}\leq \lambda,\lambda>r_{h}\right\rbrace$. Here, the horizon radius is $r_h$, and the thin-shell radius is $\lambda$. At a (2+1)-dimensional submanifold called the hypersurface, the two space-times merge, which can be represented as mathematically $\Sigma=\left\lbrace r^{\pm} =\lambda\right\rbrace$. The unique regular manifold that is produced by this method is theoretically defined as $\mathcal{M}=\mathcal{M}^{-}\cup \mathcal{M}^{+}$. The fact that $r_h<\lambda$ is ensured allows us to avoid the created structure's singularity and event horizon. The hypersurface and manifolds have the coordinates $\eta^{i}=(\tau,\theta,\phi)$ and $y^{\gamma}_\pm=(t_\pm,r_\pm,\theta_\pm,\phi_\pm)$, respectively. In this context, the proper time along the hypersurface is denoted by $\tau$. To link these coordinate systems, a coordinate transformation is employed, given as 
\begin{eqnarray}\label{7kk}
g_{ij}=\frac{\partial y^{\gamma}}{\partial\eta^{i}}\frac{\partial
y^{\beta}}{\partial\eta^{j}}g_{\gamma\beta}.
\end{eqnarray}
Also, $\Sigma:R(r,\tau)=r-\lambda(\tau)=0$ is the parametric equation that describes the hypersurface function. An examination of the physical properties of matter distribution is carried out with the help of the Lanczos equations, which are also referred to as the Einstein field equations at the hypersurface. We can do this by taking into consideration the Lanczos equations as
\begin{widetext}
\begin{eqnarray}\label{11kk}
4 \pi  \lambda\sigma&=&\sqrt{\dot{\lambda}^2+\mathcal{G}(\lambda )}-\sqrt{\dot{\lambda}^2+\mathcal{F}(\lambda )},
\\\label{12kk}
8 \pi \lambda\mathcal{P}&=&\frac{-2 \left(\dot{\lambda}^2+\ddot{\lambda} \lambda
+\mathcal{G}(\lambda )\right)-\lambda  \mathcal{G}'(\lambda )}{\sqrt{\dot{\lambda}^2+\mathcal{G}(\lambda )}}+\frac{2 \left(\dot{\lambda}^2+\ddot{\lambda}
\lambda +\mathcal{F}(\lambda )\right)+\lambda  \mathcal{F}'(\lambda )}{\sqrt{\dot{\lambda}^2+\mathcal{F}(\lambda )}},
\end{eqnarray}
\end{widetext}
here, the dash represents the derivative with respect to shell radius, and the over dot denotes the derivative with respect to $\tau$. The
the thin shell of the derived geometry is now supposed to remain
stationary at the equilibrium shell radius $\lambda_0$. Therefore,
at $\lambda=\lambda_0$, $\dot{\lambda_0}=0=\ddot{\lambda_0}$. Hence, we get
\begin{widetext}
\begin{eqnarray}\label{16kk}
4 \pi \lambda_0\sigma_0&=&\sqrt{\mathcal{G}(\lambda_0 )}-\sqrt{\mathcal{F}(\lambda_0 )}, \\\label{16kd} 8 \pi
\lambda_0\mathcal{P}_0&=&\frac{\lambda_0  \mathcal{F}'(\lambda_0 )+2
\mathcal{F}(\lambda_0 )}{\sqrt{\mathcal{F}(\lambda_0 )}}-\frac{\lambda_0  \mathcal{G}'(\lambda_0 )+2 \mathcal{G}(\lambda_0 )}{\sqrt{\mathcal{G}(\lambda_0 )}},
\end{eqnarray}
\end{widetext}
where  $\sigma_0$ is the matter density and $\mathcal{P}_0$ is the pressure of matter located at the equilibrium position.

The effective potential of the shell can be obtained from the equation of motion expressed as
$\dot{\lambda}^2+\mathcal{A}(\lambda)=0,$ which is obtained from Eq. (\ref{11kk}). Also, the potential function is named as 
$\mathcal{J}(\lambda)$ which can be expressed as
\begin{eqnarray}\label{15kk}
\mathcal{J}(\lambda)&=&-\frac{(\mathcal{G}(\lambda)-\mathcal{F}(\lambda))^2}{64 \pi ^2 \lambda ^2 \sigma
^2}+\frac{\mathcal{G}(\lambda)+\mathcal{F}(\lambda)}{2}-4 \pi ^2 \lambda ^2 \sigma ^2.~~~~~~
\end{eqnarray}
For considered cases, we get the following form of potential
function as:
\begin{itemize}

\item For the choice of CDM halo matter distribution, we get
\begin{widetext}
\begin{eqnarray}\nonumber
\mathcal{J}(\lambda)&=&\frac{1}{2} \left(-\left(\rho_s r_{s}^2
\left(\Gamma (3-\alpha ) \left(\lambda ^3 (\lambda +r_{s}) \,
_2\tilde{F}_1\left(1,4-3 \alpha ;4-\alpha ;-\frac{\lambda
}{r_{s}}\right)-r_{0}^3 (r_{s}+r_{0})
\right.\right.\right.\right.\\\nonumber&\times&\left.\left.\left.\left.\,
_2\tilde{F}_1\left(1,4-3 \alpha ;4-\alpha
;-\frac{r_{0}}{r_{s}}\right) \left(\frac{\rho_s
r_{s}^3}{\lambda  (\lambda +r_{s})^2}\right)^{-\alpha }
\left(\frac{\rho_s r_{s}^3}{r_{0}
(r_{s}+r_{0})^2}\right)^{\alpha }\right)+r_{s}
\left((1-2 \alpha ) \lambda ^3
\right.\right.\right.\right.\\\nonumber&+&\left.\left.\left.\left.r_{0}
\left(\frac{\rho_s r_{s}^3}{\lambda (\lambda
+r_{s})^2}\right)^{-\alpha }\right)\right)\right)\left(2 \alpha
\lambda ^5 (\lambda +r_{s})^2\right)^{-1}+\left(\frac{\lambda
+r_{s}}{r_{s}}\right)^{-\frac{8 \pi  \rho_s r_{s}^3}{\lambda
}}-\frac{2 m}{\lambda }+1\right)\\\nonumber&-&\left(\left(\rho_s
r_{s}^2 \left(\Gamma (3-\alpha ) \left(\lambda ^3 (\lambda
+r_{s}) \, _2\tilde{F}_1\left(1,4-3 \alpha ;4-\alpha
;-\frac{\lambda }{r_{s}}\right)-r_{0}^3
(r_{s}+r_{0})\right.\right.\right.\right.\\\nonumber&\times&\left.\left.\left.\left.
\, _2\tilde{F}_1\left(1,4-3 \alpha ;4-\alpha
;-\frac{r_{0}}{r_{s}}\right) \left(\frac{\rho_s
r_{s}^3}{\lambda  (\lambda +r_{s})^2}\right)^{-\alpha }
\left(\frac{\rho_s r_{s}^3}{r_{0}
(r_{s}+r_{0})^2}\right)^{\alpha }\right)+r_{s}
\left((1-2 \alpha ) \lambda
^3\right.\right.\right.\right.\\\nonumber&+&\left.\left.\left.\left.r_{0}
\left(\frac{\rho_s r_{s}^3}{\lambda  (\lambda
+r_{s})^2}\right)^{-\alpha }\right)\right)\right) \left(2 \alpha
\lambda ^5 (\lambda +r_{s})^2\right)^{-1}+\left(\frac{\lambda
+r_{s}}{r_{s}}\right)^{-\frac{8 \pi  \rho_s r_{s}^3}{\lambda
}}-\frac{2 m}{\lambda }-1\right){}^2\left(64 \pi ^2 \lambda ^2
\sigma ^2\right)^{-1}\\\nonumber&-&4 \pi ^2 \lambda ^2 \sigma ^2.
\end{eqnarray}
\end{widetext}

\item For the quantum wave dark matter, we have
\begin{widetext}
\begin{eqnarray}\nonumber
\mathcal{J}(\lambda)&=&-\frac{1}{28224 \pi ^2 \alpha ^2 \lambda ^4
\sigma ^2}\left(-7 \alpha  \lambda ^3 \left(\frac{\gamma  \lambda
^2}{r_{s}^2}+1\right)^{\alpha  n} \, _2F_1\left(\frac{3}{2},n
\alpha ;\frac{5}{2};-\frac{\gamma  \lambda ^2}{r_{s}^2}\right)
\left(\rho_s \left(\frac{\gamma  \lambda
^2}{r_{s}^2}+1\right)^{-n}\right)^{\alpha
}\right.\\\nonumber&+&\left.7 \alpha r_{0}^3 \left(\frac{\gamma
r_{0}^2}{r_{s}^2}+1\right)^{\alpha  n} \,
_2F_1\left(\frac{3}{2},n \alpha ;\frac{5}{2};-\frac{r_{0}^2
\gamma }{r_{s}^2}\right) \left(\rho_s \left(\frac{\gamma
r_{0}^2}{r_{s}^2}+1\right)^{-n}\right)^{\alpha }+12 \pi
\lambda  \rho_s r_{s}^2+42 \alpha  m-21 \alpha
r_{0}\right){}^2\\\nonumber&+&\frac{1}{2} \left(-\frac{1}{3
\lambda }\left(\lambda ^3 \left(\frac{\gamma \lambda
^2}{r_{s}^2}+1\right)^{\alpha  n} \, _2F_1\left(\frac{3}{2},n
\alpha ;\frac{5}{2};-\frac{\gamma  \lambda ^2}{r_{s}^2}\right)
\left(\rho_s \left(\frac{\gamma  \lambda
^2}{r_{s}^2}+1\right)^{-n}\right)^{\alpha }-r_{0}^3
\left(\frac{\gamma  r_{0}^2}{r_{s}^2}+1\right)^{\alpha
n}\right.\right.\\\nonumber&\times&\left.\left. \,
_2F_1\left(\frac{3}{2},n \alpha ;\frac{5}{2};-\frac{r_{0}^2
\gamma }{r_{s}^2}\right) \left(\rho_s \left(\frac{\gamma
r_{0}^2}{r_{s}^2}+1\right)^{-n}\right)^{\alpha }+3
r_{0}\right)-\frac{4 \pi  \rho_s r_{s}^2}{7 \alpha }-\frac{2
m}{\lambda }+2\right)-4 \pi ^2 \lambda ^2 \sigma ^2.
\end{eqnarray}
\end{widetext}

\end{itemize}

In order to explore the stability of the shell around the obtained wormhole solution, we use the conservation equation. Mathematically, we can write it as follows
\begin{equation}\label{13kk}
\frac{d}{d\tau}(4\pi
\lambda^2\sigma(\lambda))=-\mathcal{P}(\lambda) \frac{d}{d\tau}(4\pi \lambda^2),
\end{equation}
which yields
\begin{equation}\label{14kk}
\sigma'(\lambda)=-\frac{2(\sigma(\lambda)+\mathcal{P}(\sigma))}{\lambda}.
\end{equation}

Within the fields of cosmology and astrophysics, the stable configuration of thin-shell is an issue of great significance because it contributes to the investigation of potential WH solutions. In order to examine the impacts of various types of matter contents located at the hypersurface on the stable configuration of a thin shell, the equation of state (EoS) plays a crucial role. One of the many models that can describe exotic matter is a phantom-like EoS, presented as an example. It can be written as
\begin{equation}\label{17kk}
\mathcal{P}(\lambda)=\omega \sigma(\lambda),
\end{equation}
where $\omega<0$, which denotes the EoS parameter. The different ranges of
equation state parameter denotes the different types of matter
contents. It represents phantom energy, quintessence, and dark
energy state if If $\omega<-1$, $0>\omega>-1/3$ and $\omega<-1/3$,
respectively. By considering Eq.(\ref{17kk}) in (\ref{13kk}), we get
\begin{equation}\label{18kk}
\sigma'(\lambda)=-\frac{2}{\lambda}(1+\omega)\sigma(\lambda),
\end{equation}
which yields
\begin{equation}\label{19kk}
\sigma(\lambda)=\sigma(\lambda_{0})
\left(\frac{\lambda_{0}}{\lambda}\right)^{2(1+\omega)}.
\end{equation}

Now, we expand the effective potential by using the Taylor series at
equilibrium shell radius upto second-order terms as
\begin{widetext}
\begin{equation}\label{21kk}
\mathcal{J}(\lambda)=\mathcal{J}(\lambda_{0})+(\lambda-\lambda_{0})\mathcal{J}'(\lambda_{0})+\frac{1}{2}
(\lambda-\lambda_{0})^2\mathcal{J}''(\lambda_{0})+O[(\lambda-\lambda_{0})^3],
\end{equation}
\end{widetext}
where $\mathcal{J}(\lambda_{0})=0=\mathcal{J}'(\lambda_{0})$. Hence,
we have
\begin{equation}\label{22kk}
\mathcal{J}(\lambda)=\frac{1}{2}(\lambda-\lambda_{0})^2\mathcal{J}''(\lambda_{0}).
\end{equation}
It is noted that the stability of the developed structure can be
explored by using the graphical behavior of the second derivative of
effective potential at equilibrium shell radius. A geometrical
structure is said to be unstable if $\mathcal{J}''(\lambda_{0})<0$,
stable if $\mathcal{J}''(\lambda_{0})>0$ and neither stable nor
unstable if $\mathcal{J}''(\lambda_{0})=0$. Therefore, we analyze
the graphical behavior of $\mathcal{J}''(\lambda_{0})$ for CDM halo as well as quantum wave dark matter distributions.

\begin{itemize}
    \item In the framework of CDM halo matter, Figs. (\ref{f1})-(\ref{f3}) are devoted to exploring the stability of
thin-shell around WH structure for three different choices of matter EoS parameter, i.e., quintessence, dark energy, and phantom energy
type matter contents. The shell around the wormhole structure in the framework of CDM halo matter shows the stable configuration for the quintessence type distribution located at the shell see Fig.
(\ref{f1})). The stability is maximum for the smaller values of the equilibrium shell radius and decreases as the shell radius increases. It is noted that the shell stable behavior is disturbed for smaller values of wormhole throat for the choice of dark energy matter content see the left plot of Fig. (\ref{f2}). It shows the minimum stability of the shell for a higher value of wormhole throat in the right plot of Fig. (\ref{f2}). The developed structure shows maximum stable behavior for phantom energy for every choice of $\omega$ and $r_0$ see Fig. (\ref{f3}). 
\item Within the quantum wave dark matter framework, Figures (\ref{f4})-(\ref{f6}) investigate the stability of a narrow shell around a wormhole structure. The shell surrounding the wormhole structure in the quantum wave dark matter framework exhibits an unstable configuration due to the quintessence type distribution located at the shell, as shown in Figure (\ref{f4}). Stability is highest for lower equilibrium shell radii and declines as the shell radius grows with dark energy concentration, as seen in Figure (\ref{f5}). The shell stable behavior remains stable for smaller values of the wormhole throat when the dark energy matter content is chosen, as shown in the left plot of Fig. (\ref{f6}). The graph in Fig. (\ref{f6}) indicates that the stability of the shell is disrupted when the wormhole throat has a greater value. The structure becomes unstable for phantom for all values of $\omega$ and $r_0$ as seen in Fig. (\ref{f6}).
 
\end{itemize}

\section*{Concluding Remarks}

A wormhole is a fascinating speculative structure in space-time whose physical existence is still debatable. For this objective, we have tried to explore some new viable wormhole solutions under the framework of two specific energy densities via cold dark matter halo and quantum wave dark matter halo in newly developed extended $f(\mathcal{R},\mathcal{L}_m)$ gravity. In the current analysis, we have calculated field equations involving $\mathcal{L}_m=\rho$. After the embedded procedure for field equations, we have considered two specific dark matter halos as alternatives to energy density with the observational data of the M87 galaxy. The following is a list of some intriguing and feasible findings from the current study:

We have observed from the calculated shape functions that both the wormhole solutions under the effect of dark matter halos satisfy the required conditions, including flaring out and flatness conditions. All the properties for both the calculated shape functions are presented graphically in Fig. (\ref{fig1}). Further, we have calculated the valid region for the existence of the wormhole geometry under the current scenario for the involved model parameter $\alpha$ with specific range $-8\leq\alpha\leq8$, which is reported in Fig. (\ref{fig2}). 

Overall, research into wormhole geometry and associated energy conditions has a critical role in understanding the theoretical and physical consequences of wormhole space-time structures and establishing the concept that wormholes may exist in the universe. In the current analysis, all the energy conditions have valid and invalid regions for the different ranges of involved model parameter $\alpha$. The energy density with $-8\leq\alpha\leq8$ remains positive under the effect of two different kinds of dark matter halos. The valid region for energy density has been confirmed from Fig. (\ref{fig3}). The NEC has strongly violated for $-8\leq\alpha\leq8$ due to the maximum invalid region, which has been confirmed from Fig. (\ref{fig4}). The NEC violation has a positive indication of the presence of exotic matter. The presence of exotic matter allows the wormhole to exist. All the other energy conditions, including SEC, are presented in Figs. (\ref{fig5}-\ref{fig10}) for $-8\leq\alpha\leq8$.

Figs. (\ref{f1})-(\ref{f3}) investigate the stability of thin-shell around a wormhole structure with quintessence, dark energy, and phantom energy matter contents in the setting of CDM halo matter. The shell stability is maximum for quintessence, diminishes with increasing shell radius, and becomes unstable for smaller wormhole throats containing dark energy. It is noted that the developed structure exhibits maximal stability for phantom energy for suitable values of $\omega$ and $r_0$. Figures (\ref{f4})-(\ref{f6}) in the quantum wave dark matter framework study the stability of a tight shell around a wormhole structure. Figure (\ref{f4}) depicts an unstable shell configuration around the wormhole structure in the quantum wave dark matter framework caused by the quintessence type distribution. Figure (\ref{f5}) shows that stability is most compelling at lower equilibrium shell radii and decreases with increasing dark energy concentration. The shell stable behavior stays stable for smaller values of the wormhole throat when dark energy matter content is chosen, as illustrated in the left plot of Fig. (\ref{f5}). The graph in Fig. (\ref{f5}) shows that a larger wormhole throat disrupts the shell's stability, and the developed structure becomes unstable for phantom energy (\ref{f6}).


\section*{References}
\begin{thebibliography}{}

\bibitem{misnerwheeler} C. W. Misner and J. A. Wheeler, Annals Phys. 1957, 2, 525.
\bibitem{weyl1} H. Weyl, AnnaLen Der Physik, 1921, 370, 541.
\bibitem{einsteinrosen} A. Einstein and N. Rosen, Phys. Rev. 1935, 48 73.
\bibitem{EHT1} EHT Collaboration et al., First M87 event horizon telescope results. I. The
shadow of the supermassive black hole, ApJ, 2019c, 875, L1.
\bibitem{EHT2} EHT Collaboration et al., ApJL, 2019b, 875, L5. 
\bibitem{EHT3} EHT Collaboration et al., ApJL, 2019c, 875, L6. 
\bibitem{mt1} M.S. Morris and K.S. Thorne, Amer. J. Phys., 1988, 56, 395.
\bibitem{gw} B. P. Abbott et. al,Phys. Rev. Lett., 2016, 116, 6.
\bibitem{shaikh} R. Shaikh, P. Banerjee, S. Paul and T. Sarkar, JCAP, 2019, 028. 

\bibitem{njim1} S. Perlmutter, et al., Bull. Am. Astron. Soc. 29 (1997) 1351, 
\bibitem{njim2} A.G. Riess, et al., Astron. J. 116 (1998) 1009-1038, 
\bibitem{njim3} S.M. Carroll, V. Duvvuri, M. Trodden, M.S. Turner, Phys. Rev. D 70 (2004) 043528,

\bibitem{njim4} S. Capozziello, V.F. Cardone, A. Troisi, 
Mon. Not. Roy. Astron. Soc. 375 (2007) 1423-1440, 
\bibitem{njim5} S. Nojiri, S.D. Odintsov, Phys. Lett. B 657 (2007)
238-245.
\bibitem{rjim1} T. Harko, F.S.N. Lobo, $f(R, L_m)$ gravity Eur. Phys. J. C 70 (2010) 373–379, http://dx.doi.org/10.1140/epjc/s10052-010-1467-3.
\bibitem{rjim2} O. Bertolami, F.S.N. Lobo, J. Pàramos, Phys. Rev. D 78 (2008) 064036.
\bibitem{rjim3} O. Bertolami, C.G. Boehmer, T. Harko, et al., Phys. Rev. D 75 (2007) 104016.
\bibitem{rjim4} J. Wang, K. Liao, Classical Quantum Gravity 29 (2012) 215016.
\bibitem{rjim5} L.V. Jaybhaye, R. Solanki, S. Mandal, et al., Phys. Lett. B 831 (2022) 137148.

\bibitem{njim10} S. Capozziello and M. De Laurentis, Phys. Rep. 509
(2011) 167.
\bibitem{njim11} S. Capozziello and M. Francaviglia, Gen. Relativ. Gravit. 40 (2008) 357.
\bibitem{njim12} S. Capozziello et al., Phys. Lett. B 639 (2006) 135.
\bibitem{njim13} S. Capozziello, V. F. Cardone and A. Troisi, Phys. Rev. D 71 (2005) 043503.
\bibitem{njim14} S. Capozziello, A. Stabile and A. Troisi, Class. Quantum Grav. (2007) 2153.
\bibitem{njim15} S. Capozziello et al., Phys. Rev. D 83 (2011) 064004.
\bibitem{njim16} S. Capozziello et al., Phys. Rev. D 86 (2012) 127504.
\bibitem{njim17} S. Capozziell, R. Pincak, E. N. Saridakis et al., Annals Phys. 390 (2018) 303.
\bibitem{bohmer2012} C.G. B\"ohmer, T. Harko and  F.S.N. Lobo,  Phys. Rev.D  85, 044033 (2012)
\bibitem{lobo2009} F.S.N. Lobo and M.A. Oliveira, Phys. Rev. D 80, 104012 (2009)
\bibitem{lobo2020}  F.S.N. Lobo, A. Simpson and M. Visser,  Phys. Rev. D 101, 124035 (2020)
\bibitem{harko2013} T. Harko, F.S.N. Lobo, M.K. Mak et al.,  Phys. Rev. D 87, 067504 (2013)
\bibitem{kanti2011} 	P. Kanti, B. Kleihaus and J. Kunz, Phys. Rev. Lett. 107, 271101 (2011)
\bibitem{usmani2010} A.A. Usmani,  Z. Hasan, F. Rahaman, Gen. Relativ. Gravit. 42, 2901 (2010)
\bibitem{rahaman2006} F. Rahaman, M Kalam, M Sarker, et al., Phys. Lett. B 633, 2-3 (2006)
\bibitem{rahaman2012} F. Rahaman, S. Islam, P. K. F. Kuhfittig, et al.,  Phys. Rev. D 86, 106010 (2012)
\bibitem{zubair2016} M. Zubair, S. Waheed and Y. Ahmad, Eur. Phys. J. C 76, 444 (2016)
\bibitem{ovgun2018} 	A. \"{O}vg\"{u}n, Phys. Rev. D 98, 044033 (2018)
\bibitem{mustafa2021}  G. Mustafa, M. Ahmad, A. \"Ovg\"un, et al., Fortschr. Phys. \textbf{69} 2100048 (2021)

\bibitem{cap2} S. Capozziello, R. Pincak and E. Bartos, Symmetry 12, 774 (2020)
\bibitem{cap3} S. Capozziello, O. Luongo and L. Mauro, Eur. Phys. J. Plus 136, 167 (2021)
\bibitem{cap4} S. Capozziello and Nisha Godani, Phys. Lett. B 835, 137572  (2022)
\bibitem{cap5} S. Capozziello, T. Harko, T.S. Koivisto et al., Physical Review D 86, 127504 (2012)
\bibitem{kuhfittig2015} P.K.F. Kuhfittig, Eur. Phys. J. C 74 2818 (2014)
\bibitem{kuhfittig2005} P.K.F. Kuhfittig, Phys. Rev. D 73, 084014 (2006)
\bibitem{ref1e}  Javed, F., and Ji Lin. "Novel gravastar solutions: Investigating stability, energy, and entropy in the presence of cloud of strings and quintessence." Chinese Journal of Physics (2024).
\bibitem{ref1e2}  Javed, F.: Annals of Physics 458 (2023): 169464.
\bibitem{ref1e3} Sharif, M., and F. Javed.: Journal of Experimental and Theoretical Physics 133.4 (2021): 439-448.
\bibitem{ref1e4} Sharif, M., and F. Javed.: Astronomy Reports 65.5 (2021): 353-361.

\bibitem{ref1} V. De Falco, E. Battista, S. Capozziello et al, Eur. Phys. J. C 81, 1 (2021)

\bibitem{njim18} N.S. Kavya et al., Chinese Journal of Physics 87 (2024) 751–765
\bibitem{njim19} N.S. Kavya et al., Annals of Physics 455 (2023) 169383.
\bibitem{njim20} N.S. Kavya et al., Chinese Journal of Physics 84 (2023) 1–11


\bibitem{ta10} F. Javed, G. Mustafa, A. OvgUn, and M. F. Shamir.: Euro. Phys. J. Plus 137, 1-16 (2022)

\bibitem{ta11} F. Javed, S. Mumtaz, G. Mustafa, I. Hussain, and Wu-Ming Liu.: Euro. Phys. J. C 82, 825 (2022)

\bibitem{ta12} F. Javed, S. Sadiq, G. Mustafa, and I. Hussain.:  Physica Scripta 97, 125010 (2022)

\bibitem{ta13}G. Mustafa, X. Gao, and F. Javed.: Fortschritte der Physik 70, 2200053 (2022)

\bibitem{ta14}G. Mustafa,  S. K. Maurya, S. Ray, and F. Javed.: Annals of Physics 169551 (2023)




\bibitem{DM1} H.-Y. Schive, T. Chiueh, T. Broadhurst, Nature Phys. 2014, 10, 496, 1406.6586.
\bibitem{DM3} A. Herrera-Martín, M. Hendry, A. X. Gonzalez-Morales, L. A. Ureña López, Astrophys. J. 2019, 872, 11, 1707.09929.
\bibitem{DM2} H.-Y. Schive, M.-H. Liao, T.-P. Woo, S.-K. Wong, T. Chiueh, T. Broadhurst, W. Y. P. Hwang, Phys. Rev. Lett. 2014, 113, 261302, 1407. 7762.

\bibitem{DM4} P. K. F. Kuhfittig, Eur. Phys. J. C 74, 2818 (2014), arXiv:1311.2274 [gr-qc] .
\bibitem{DM5} F. Rahaman, P. K. F. Kuhfittig, S. Ray, and N. Islam, Eur. Phys. J. C 74, 2750 (2014), arXiv:1307.1237 [gr-qc] .
\bibitem{DM6} J. F. Navarro, C. S. Frenk, S. D. M. White, A universal density profile from hierarchical clustering. ApJ, 490, No. 2, 493 (1997).
\bibitem{DM7} L. J. Oldham and M. W. Auger, Galaxy structure from multiple tracers  II. M87 from parsec to megaparsec scales, Mon. Not. R. Astron. Soc. 457, 421 (2016).

\bibitem{th1}  Zhao, Yuqian, et al. arXiv preprint arXiv:2303.09215 (2023).
\bibitem{th2} Liu, Dong, et al. The European Physical Journal C 83.7 (2023): 565.
\bibitem{th3} Naoz, Smadar, Joseph Silk, and Jeremy D. Schnittman. The Astrophysical Journal Letters 885.2 (2019): L35.
\bibitem{th4} Macedo, Caio FB, et al. The Astrophysical Journal 774.1 (2013): 48.
\bibitem{th5}  Zhang, Chao, et al. Physics of the Dark Universe 37 (2022): 101078.
\bibitem{fg1} Pantig, Reggie C., and Ali Övgün. Fortschritte der Physik 71.1 (2023): 2200164.
\bibitem{fg2} Kawai, Hikaru, and Yuki Yokokura.  Universe 6.6 (2020): 77.





\end{thebibliography}
\end{document}